\documentclass[11pt]{article}

\usepackage[usenames,dvipsnames]{xcolor}
\usepackage[a4paper]{geometry}
\usepackage[english]{babel}
\usepackage{cite,enumerate,enumitem,booktabs,float,graphicx}

\usepackage[T1]{fontenc}
\usepackage[utf8]{inputenc}
\usepackage{lmodern}

\makeatletter
\g@addto@macro\bfseries{\boldmath}
\makeatother

\usepackage{bm,amsmath,amssymb,tensor,mathtools}
\numberwithin{equation}{section}
\usepackage{nth}

\usepackage{tikz}
\usetikzlibrary{cd}

\usepackage{caption}
\usepackage{subcaption}

\usepackage[pdftex]{hyperref}
\definecolor{dark-blue}{rgb}{0.15,0.15,0.4}
\hypersetup{
  colorlinks,
  linkcolor={dark-blue},
  citecolor={blue}
}

\newcommand{\pd}{\partial}

\newcommand{\spa}{\ , \ \ }
\newcommand{\CL}{\mathcal{L}}

\usepackage{enumitem}
\usepackage{amsthm}

\usepackage{authblk}

\definecolor{darkgreen}{rgb}{0.0, 0.4, 0.1}
\definecolor{darkyellow}{rgb}{0.75, 0.75, 0.1}

\title{\textbf{%
  Longitudinal Galilean and Carrollian limits
  \\ of non-relativistic strings}
}
\date{}
\author[1]{Leo Bidussi}
\author[2]{Troels Harmark}
\author[1]{Jelle Hartong}
\author[2,3]{Niels A. Obers}
\author[3]{\protect\\ and Gerben Oling}
\affil[1]{%
  School of Mathematics and Maxwell Institute for Mathematical Sciences,
  \protect\\
  University of Edinburgh,
  Peter Guthrie Tait Road,
  Edinburgh EH9 3FD, UK
}
\affil[2]{%
  Niels Bohr International Academy, Niels Bohr Institute,
  University of Copenhagen,
  \protect\\
  Blegdamsvej 17,
  DK-2100 Copenhagen Ø,
  Denmark
}
\affil[3]{%
  Nordita,
  KTH Royal Institute of Technology and Stockholm University,
  \protect\\
  Hannes Alfvéns väg 12,
  SE-106 91 Stockholm,
  Sweden
}

\begin{document}

\maketitle
\thispagestyle{empty}

\begin{abstract}
\noindent
It is well known that one can take an infinite speed of light limit that gives rise to non-relativistic strings with a relativistic worldsheet sigma model but with a non-relativistic target space geometry. In this work we systematically explore two further limits in which the worldsheet becomes non-Lorentzian. The first gives rise to a Galilean string with a Galilean structure on the worldsheet, extending previous work on Spin Matrix-related string theory limits. The second is a completely novel limit leading to a worldsheet theory with a Carrollian structure. 
We find the Nambu--Goto and Polyakov formulations of both limits and explore gauge fixing choices. Furthermore, we study in detail the case of the Galilean string for a class of target space geometries that are related to Spin Matrix target space geometries, for which the Nambu--Goto action (in static gauge) is quadratic in the fields.
\end{abstract}

\newpage
\tableofcontents

\section{Introduction}

The development of non-relativistic (NR) string theory in recent years has shown that it provides a fertile arena to study 
quantum gravity, holography and novel classes of quantum field theories. 
First of all,
there has been significant progress in formulating the flat space NR string theory
of Gomis and Ooguri \cite{Gomis:2000bd} (see also \cite{Danielsson:2000gi})
on curved NR backgrounds~\cite{Harmark:2017rpg,Kluson:2018egd,Bergshoeff:2018yvt,Harmark:2018cdl,Bergshoeff:2018vfn,Harmark:2019upf,Bergshoeff:2021bmc,Yan:2021lbe,Bidussi:2021ujm}. 
This built in part on advances in understanding Newton-Cartan (NC) geometry~\cite{Andringa:2010it} and its stringy analogues \cite{Andringa:2012uz}, as well as the introduction of torsional Newton-Cartan geometry \cite{Christensen:2013lma,Christensen:2013rfa,Hartong:2015wxa}, culminating in our present understanding of
string torsional string Newton-Cartan (TSNC) geometry \cite{Bidussi:2021ujm} (see also~\cite{Bergshoeff:2021bmc}). 
The resulting string theories appear to comprise a self-consistent UV complete corner of string theory with a non-relativistic spectrum. In particular, using the fact that the worldsheet theories are 
still two-dimensional relativistic conformal field theories,  beta functions have been obtained in 
\cite{Gomis:2019zyu,Gallegos:2019icg,Yan:2019xsf,Bergshoeff:2019pij}.
See \cite{Oling:2022fft} for a recent review of NR  string theory.

An important second strand of investigation has been the search for a non-relativistic version~\cite{Harmark:2017rpg,Harmark:2018cdl,Harmark:2019upf,Harmark:2020vll} of the AdS/CFT correspondence.%
\footnote{Related work on NR strings in AdS backgrounds was done in  
Refs.~\cite{Gomis:2005pg,Roychowdhury:2020kma,Roychowdhury:2020yun,Fontanella:2021hcb,Roychowdhury:2021wte,Fontanella:2021btt,Kluson:2021tub,Fontanella:2022pbm,Fontanella:2022wfj}.} 
This has revealed a further class of NR string theories described by non-relativistic worldsheet models. In particular, 
while the TSNC string still has a Lorentzian worldsheet structure but a non-Lorentzian target spacetime, these new worldsheet models have non-Lorentzian geometry on both the worldsheet as well as the target spacetime.  More specifically, 
the sigma models with a Galilean structure on the worldsheet that have been found in \cite{Harmark:2017rpg,Harmark:2018cdl,Harmark:2019upf,Harmark:2020vll} are believed to correspond to strings that are dual to Spin Matrix theory limits of $\mathcal{N}=4$ SYM \cite{Harmark:2014mpa}.
In that case, the target spacetime was identified
as a novel type of non-Lorentzian geometry dubbed $U(1)$-Galilean geometry.

In the first part of this paper (Section \ref{sec:limits}),  we will considerably extend the class of sigma models that could correspond to string theories with non-Lorentzian worldsheet geometries.
Building on the recent TSNC formulation of NR string theory in~\cite{Bidussi:2021ujm}, we find  two new classes of sigma models, arising from scaling limits of the TSNC string.
One class that we name the {\sl Galilean string} has a Galilean structure on the worldsheet, and a new non-Lorentzian target-space geometry which generalizes the aforementioned $U(1)$-Galilean geometry. 
The other class that we name the {\sl Carrollian string} has a Carrollian structure on the worldsheet, and yet another type of non-Lorentzian target-space geometry.%
\footnote{%
  We use these names for brevity, but note that other strings with Galilean and Carrollian worldsheet structures have appeared from distinct limits in previous literature, see for example~\cite{Isberg:1993av,Bagchi:2020fpr,Batlle:2016iel}.
}
For both the Galilean and Carrollian strings, we find a Nambu--Goto type as well as a Polyakov type sigma model. We note that the target-space geometry in both cases inherits the two-form of TSNC geometry which couples to the tension current of the string. We emphasize that the two theories exhibit a longitudinal Galilean and Carrollian boost symmetry respectively. Furthermore, we show that the Galilean and Carrollian string sigma models can be mapped to each other by exchanging the time and space longitudinal vielbeine.

In the second part of this paper (Sections \ref{sec:ws_gauge_fixings} and \ref{sec:SMT_target}) we examine more closely the new sigma models found from the longitudinal limits. This is done with the aim of taking preliminary steps towards a quantization of these new worldsheet theories.

Firstly, we consider simplifying gauge fixings of the worldsheet diffeomorphisms of the sigma models. We develop a version of the static gauge for the Nambu--Goto type sigma models, which is particularly useful in these settings, since both the Galilean and Carrollian string Nambu--Goto sigma models do not have the square root that remains after this gauge fixing in the usual relativistic string. 
We also find a related but distinct partial gauge fixing for the Galilean string, which eliminates certain negative-definite kinetic terms in both the Nambu--Goto and Polyakov sigma models.

Secondly, we exhibit a family of target space geometries as a solvable model for the Galilean string. Backgrounds of this type are known to arise in the context of the Spin Matrix limit of the AdS/CFT correspondence \cite{Harmark:2017rpg,Harmark:2018cdl,Harmark:2019upf,Harmark:2020vll}.
We analyze their phase space description and find the space of solutions.
Furthermore, we exhibit the global symmetries of these backgrounds by determining the Noether charges.

Finally, in Section \ref{sec:concl} we present our conclusions and give an outlook. 

For completeness, we present in Appendix \ref{app:SymAlg} the underlying local symmetry algebra of the novel target space geometry of both the Galilean and Carrollian string.
This is obtained from an appropriate İnönü--Wigner contraction of the fundamental string Galilean algebra of the TSNC string~\cite{Bidussi:2021ujm}.

\section{Longitudinal limits of non-relativistic string theory}
\label{sec:limits}

In this section we introduce two distinct limits of the Torsional String Newton-Cartan (TSNC) string \cite{Bidussi:2021ujm,Gomis:2000bd,Danielsson:2000gi,Harmark:2019upf,Bergshoeff:2019pij,Bergshoeff:2021tfn,Oling:2022fft}. 
As we briefly review in Section~\ref{sec:TSNC_review}, the TSNC string arises from relativistic string theory by taking an infinite speed of light limit along the directions transverse to the string worldsheet. 
The target geometry of the TSNC string has a (1+1)-dimensional light cone structure associated with the longitudinal directions that lie along the string worldsheet.
There is thus a longitudinal speed of light in the TSNC target space that we can send to infinity or to zero, as we emphasize in Section~\ref{sec:long_speed_light}.

The first limit, which we introduce in Section~\ref{sec:Gal_long_limit}, is a Galilean limit that sends the longitudinal speed of light to infinity.
It generalizes previous limits obtained and studied in Refs.~\cite{Harmark:2017rpg,Harmark:2018cdl,Harmark:2019upf,Harmark:2020vll} for a less general class of target spacetimes, and has sometimes been referred to as the Spin Matrix Theory (SMT) string.  
The second limit, in Section~\ref{sec:Car_long_limit}, is completely novel.
This is a Carrollian limit, which sends the longitudinal speed of light of the TSNC string to zero. In Sections~\ref{sec:Gal_long_limit} and~\ref{sec:Car_long_limit}, the limits are taken of the TSNC Nambu--Goto action, but in Section~\ref{sec:limits_Pol} we describe the equivalent limits for the TSNC Polyakov action.

\subsection{Brief review of non-relativistic string theory}
\label{sec:TSNC_review}

We begin with a brief review of the limit of the NS-NS sector of relativistic string theory (omitting the dilaton) in which one takes the speed of light along the transverse directions to infinity, following \cite{Bidussi:2021ujm} (see also~\cite{Bergshoeff:2021bmc} and more generally~\cite{Gomis:2000bd,Danielsson:2000gi,Harmark:2019upf,Bergshoeff:2019pij,Bergshoeff:2021tfn,Oling:2022fft}). 
Consider the Nambu--Goto (NG) action for a relativistic closed bosonic string in a target spacetime with a metric $g_{MN}$ and Kalb--Ramond 2-form $B_{MN}$,
\begin{equation}
\label{relNG}
    S=-c \, T_F \int d^2\sigma\left(\sqrt{-\text{det}(g_{\alpha\beta})}+\frac{1}{2}\epsilon^{\alpha\beta}B_{\alpha\beta}\right)\,.
\end{equation}
Here, $c$ is the speed of light and 
$T_{\rm F}$ is the tension (in units of mass per unit length).
The $\sigma^\alpha= (\tau,\sigma)$ are the worldsheet coordinates.
Furthermore, $g_{\alpha\beta}$ is the pullback of the target space-time metric $g_{MN}$ with respect to the embedding coordinates $X^M(\sigma^\alpha)$, and $B_{\alpha\beta}$ is the pullback of the Kalb--Ramond field $B_{MN}$.
The $M,N=0,1,...,d+1$ are target space-time indices with $d+2$ being the space-time dimension. The 2-dimensional epsilon tensor is defined such that $\epsilon^{\tau\sigma}=-\epsilon_{\tau\sigma}=+1$. 
We denote the values of the $\alpha, \beta$ indices by $\tau, \sigma$ as opposed to $0$ and $1$ to avoid confusion with other indices.

To obtain what is usually referred to as non-relativistic string theory, we then consider a $c\rightarrow \infty$ limit of strings with a non-zero winding along a direction which also carries a near-critical Kalb--Ramond field flux. 
Specifically, we assume that one can expand the metric and Kalb--Ramond field for large $c$ as~\cite{Bidussi:2021ujm}%
\footnote{%
  Note that $\tau_M^A$ here corresponds to
  $\tau^A_M - \frac{1}{2} c^{-2} \pi^B_M \epsilon_{BC} \eta^{CA}$
  in~\cite{Bidussi:2021ujm}.
}
\begin{equation}
\begin{array}{c}
\label{eq:gB_largec}
g_{MN} = c^2 \eta_{AB} \tau^A_M \tau^B_N  + h_{MN} + \mathcal{O} (c^{-2})\,,
\\[2mm]
B_{MN} = c^2 \epsilon_{AB} \tau^A_M \tau^B_N + m_{MN} + \mathcal{O} (c^{-2})\,.
\end{array}
\end{equation}
Here, we have introduced a pair of vielbeine $\tau^A_M$, with $A=0,1$, which delineate the two directions longitudinally along the string worldsheet, one of which is timelike and one of which is spacelike and compact.
Additionally, we have a transverse tensor $h_{MN}$ with signature $(0,0,1,\ldots,1)$, and a two-form $m_{MN}$, which enters through the subleading part of the Kalb--Ramond field.
Furthermore, $\eta_{AB}=\mbox{diag}(-1,1)$ is the two-dimensional Minkowski metric in the longitudinal directions.

At this point, it is useful to pause and consider some of the dimensional consequences of the parametrization~\eqref{eq:gB_largec} above.
Recall that the line element $ds^2=g_{MN}dx^M dx^N$ has dimensions of length squared.
In this paper, we will use a time coordinate~$t=x^0$ in target space with dimensions of time, while the other coordinates $x^1$ to $x^{d+1}$ have dimensions of length.%
\footnote{%
  Since we will often encounter Weyl invariance on the worldsheet, we take the worldsheet coordinates to be dimensionless.
}
Introducing $c$ as in the expansion~\eqref{eq:gB_largec} above then separates the local propagation speed of light along the longitudinal directions
(corresponding to $\tau^0$ and $\tau^1$)
from the propagation speed along the transverse spatial directions. 
To see this, suppose that for a given event one goes to a local rest frame with 
$\tau^0 = dt$, $\tau^1 = \tilde{c}^{-1} dx^1$
and $h_{MN} dx^M dx^N = \sum_{i=1}^d (dx^{1+i})^2$.
Here, we have to introduce a velocity $\tilde{c}$ since $\tau^1$ is in units of time due to~\eqref{eq:gB_largec} while~$x^1$ is in units of space.
Then the speed of light in the $x^1$ direction is $\tilde{c}$, while it is $c$ in the transverse directions $x^M$, $M=2,3,...,d+1$.
Furthermore, note that we cannot remove~$\tilde{c}$ by rescaling the $x^1$ coordinate, since this direction is periodic and the string worldsheet winds around it.

Taking $c$ as in~\eqref{eq:TSNCstringwithc} to be large means that the speed of light along the transverse directions is much larger than in the longitudinal directions, which results in an emerging stringy Galilean causal structure in target space.%
\footnote{%
    Note that when taking limits with respect to $c$, or below with respect to $\tilde{c}$, they should be understood with respect to the ratio of a characteristic velocity to these respective velocities.
}
However, a two-dimensional Lorentzian structure along the longitudinal directions remains, with associated metric
\begin{equation}
\label{longi_met}
\tau_{MN} = \eta_{AB} \tau^A_M \tau^B_N = - \tau^0_M \tau^0_N + \tau^1_M \tau^1_N\,.
\end{equation}
As we will see below, this induces a Lorentzian structure on the string worldsheet.
One of the main goals of this work is to systematically obtain the actions that result from Galilean and Carrollian limits of this longitudinal Lorentz structure.

First, let us continue with the large transverse speed of light limit.
The leading (divergent) part of the metric and Kalb--Ramond field in~\eqref{eq:gB_largec} are tuned such that any possible divergent terms in \eqref{relNG} cancel in the $c\rightarrow \infty$ limit,
where one obtains \cite{Bidussi:2021ujm}
\begin{equation}
\label{NG_TSNC}
S = - \frac{\mathcal{T}}{2} \int d^2 \sigma \Big[ \sqrt{-\tau}\, \eta^{AB} \tau_A^\alpha \tau_B^\beta h_{\alpha\beta} + \epsilon^{\alpha\beta} m_{\alpha\beta} \Big] \,,
\end{equation}
with $\mathcal{T} = c\, T_{\rm F}$ fixed in the limit.
The longitudinal Lorentz structure $\tau_M^A$ pulls back to a two-dimensional Lorentzian structure $\tau_\alpha^A$ on the worldsheet, whose determinant and inverse vielbeine are given by
\begin{equation}
  \sqrt{-\tau}
  = \epsilon^{\alpha\beta} \tau^0_\alpha \tau^1_\beta
  \spa
  \tau_A^\alpha
  = - \frac{\epsilon^{\alpha\beta}\epsilon_{AB} \tau^B_\beta}{\sqrt{-\tau}}\,.
\end{equation}
In addition to the longitudinal directions, the string couples to the transverse directions through the pullback of $h_{MN}$ and also to the pullback of the two-form $m_{MN}$.
The corresponding target space-time geometry is known as torsional string Newton-Cartan (TSNC) geometry, since its natural connections can in general be torsionful, and we refer to the formulation of non-relativistic string theory as the TSNC string.

Now let us consider the local symmetries of the Nambu--Goto type action~\eqref{NG_TSNC} for the TSNC string.
For this, it is useful to introduce the transverse vielbeine $e^a_M$ by $h_{MN}=\delta_{ab}e^a_M e^b_N$ where $a,b=2,\ldots,d+1$.
It follows from the above limit that $(\tau^A_M, e^a_M)$ should be an invertible $(d+2)\times (d+2)$ matrix.
By inverting this matrix one can introduce the inverse longitudinal and transverse vielbeine $\tau_A^M$ and $e_a^M$, and the inverse transverse metric $h^{MN}=\delta^{ab} e_a^M e_b^N$, such that%
\footnote{Note that we raise and lower longitudinal vielbein indices $A,B$ with $\eta^{AB}$ and $\eta_{AB}$, respectively, while the transverse vielbein indices $a,b$ are raised and lowered with a Kronecker delta.}
\begin{equation}
\tau_A^M e^a_M = 0 \spa \tau^A_M e_a^M = 0 \spa \tau_A^M \tau^B_M = \delta_A^B \spa e_a^M e^b_M = \delta_a^b
\end{equation}
\begin{equation}
h_{MP} h^{PN} + \tau^A_M \tau_A^N = \delta_M^N\,.
\end{equation}
We then find that the local symmetries of the action~\eqref{NG_TSNC} are:
\begin{itemize}

\item Worldsheet diffeomorphisms $\delta\sigma^\alpha=-\zeta^\alpha$ that act on the embedding scalars as 
\begin{equation}
\label{SYM_WSdiff}
\delta X^M=\zeta^\alpha\partial_\alpha X^M\,.
\end{equation}

\item One-form gauge transformations of the two-form
\begin{equation}
\label{SYM_twoform}
\delta m_{MN} = \partial_M \Sigma_N - \partial_N \Sigma_M\,.
\end{equation}
\item Weyl transformations of the longitudinal vielbeine
\begin{equation}
\label{SYM_Weyl}
    \delta\tau^A_M=\Omega \tau^A_M\,,
\end{equation}
with $\Omega(x^M)$ an arbitrary function on the target space. 
\item Local rotations, where the $e^a_M$ transform as a vector under local $SO(d)$ rotations with $a$ a $d$-dimensional rotation index.
\item Local Lorentz boosts, where the $\tau^A_M$ transform in a vector representation of $SO(1,1)$ with $A$ a two-dimensional Lorentz index.
\item Local string Galilean boost transformations with parameter $\lambda^A_{a}$
\begin{equation}
\label{stringGal}
\delta h_{MN} = - \eta_{AB}\lambda^B_{b} ( \tau^A_M e^b_N + \tau^A_N e^b_M )
\spa
\delta m_{MN} = -  \epsilon_{AB} \lambda^B_b ( \tau^A_M e^b_N - \tau^A_N e^b_M )\,.
\end{equation}
\end{itemize}
We emphasize that the Weyl transformations are very different from the ones in ordinary relativistic string theory where they are associated with the worldsheet metric of the Polyakov action.
While such worldsheet Weyl transformations will also appear in Section~\ref{sec:limits_Pol} below, the ones above act on target space fields instead.
See Appendix~\ref{app:SymAlg} for further details on the underlying symmetry algebra of TSNC geometry.

For certain target spaces the NG action \eqref{NG_TSNC} can also have global symmetries. When that is the case these are
general coordinate transformations in target space, generated infinitesimally by $\delta X^M=-\xi^M$. Demanding that the NG action \eqref{NG_TSNC} is invariant under such transformations leads to the condition
\begin{eqnarray}
    0 & = & \frac{cT_F}{2}\int d^2\sigma\left[\sqrt{-\tau}\tau^{\alpha\beta}\partial_\alpha X^M\partial_\beta X^N\mathcal{L}_\xi h_{MN}-\frac{1}{2}h_{\alpha'\beta'}G^{\alpha'\beta'\alpha\beta}\partial_{\alpha}X^M\partial_\beta X^N\mathcal{L}_\xi\tau_{MN}\right.\nonumber\\
    &&+\epsilon^{\alpha\beta}\partial_\alpha X^M\partial_\beta X^N\mathcal{L}_\xi m_{MN}\left.\right]\,,\label{eq:globalsym}
\end{eqnarray}
where we defined
\begin{equation}
    \delta_X\left(\sqrt{-\tau}\tau^{\alpha'\beta'}\right)=\frac{1}{2}G^{\alpha'\beta'\alpha\beta}\partial_{\alpha}X^M\partial_\beta X^N\mathcal{L}_\xi\tau_{MN}\,,
\end{equation}
with
\begin{equation}
    G^{\alpha'\beta'\alpha\beta}=\tau^{\alpha'\alpha}\tau^{\beta'\beta}+\tau^{\alpha'\beta}\tau^{\beta'\alpha}-\tau^{\alpha'\beta'}\tau^{\alpha\beta}\,,
\end{equation}
being the two-dimensional DeWitt metric.
Using the fact that the target space objects $\tau_{MN}$ etc.\
are defined up to the local transformation given above, requiring \eqref{eq:globalsym} is equivalent to demanding that
\begin{eqnarray}
0=\delta \tau^A_M & := & \mathcal{L}_\xi\tau^A_M+\Omega\tau_M^A+\lambda\epsilon^A{}_B\tau^B_M\,,\\    
0=\delta h_{MN} & := & \mathcal{L}_\xi h_{MN}- \eta_{AB}\lambda^B_{b} ( \tau^A_M e^b_N + \tau^A_N e^b_M )\,,\\    
0=\delta m_{MN} & := & \mathcal{L}_\xi m_{MN}-  \epsilon_{AB} \lambda^B_b ( \tau^A_M e^b_N - \tau^A_N e^b_M )+\partial_M \Sigma_N - \partial_N \Sigma_M\,,
\end{eqnarray}
where $\lambda$ is an infinitesimal local Lorentz transformation.

\subsection{Introducing the longitudinal speed of light}
\label{sec:long_speed_light}

Following the discussion after Equation~\eqref{eq:gB_largec} above, we now introduce a longitudinal speed of light to enable us to consider Galilean and Carrollian limits in the longitudinal directions below. 
Since the $\tau^A_Mdx^M$ are both in units of time above,
we now make the replacement
\begin{equation}
\label{new_tau1}
    \tau^1_M \rightarrow \frac{1}{\tilde{c}} \tau^1_M\,,
\end{equation}
where $\tilde{c}$ is the longitudinal speed of light as explained above.
After this rescaling, $\tau^1_Mdx^M$ is in units of space while $\tau^0_Mdx^M$ is still in units of time.
The longitudinal Minkowski metric is then $\eta_{AB}= \mbox{diag}(-1,\tilde{c}^{-2})$ so that the longitudinal metric \eqref{longi_met} becomes
\begin{equation}
\tau_{MN} = - \tau^0_M \tau^0_N + \frac{1}{\tilde{c}^2} \tau^1_M \tau^1_N\,.
\end{equation}
The TSNC NG action \eqref{NG_TSNC} written with explicit factors of $\tilde{c}$ is then
\begin{equation}\label{eq:TSNCstringwithc}
S =  \frac{\mathcal{T}}{2\tilde{c}}\int d^2 \sigma \left[  \sqrt{-\tau}\left( \tau_0^\alpha \tau_0^\beta-\tilde c^2\tau_1^\alpha \tau_1^\beta\right) h_{\alpha\beta} -  \epsilon^{\alpha\beta} m_{\alpha\beta} \right] \,,
\end{equation}
where we made the additional replacement $m_{MN} \rightarrow \tilde{c}^{-1} m_{MN}$.
For future purposes, we also note that the local string Galilean boost transformations now take the form 
\begin{equation}
\label{stringGal_tildec}
\begin{array}{c}
\delta h_{MN} = \lambda^0_{b} ( \tau^0_M e^b_N + \tau^0_N e^b_M ) -  \lambda^1_{b} ( \tau^1_M e^b_N + \tau^1_N e^b_M )
\\[2mm]
 \delta m_{MN} =  \tilde{c}^2\lambda^1_b ( \tau^0_M e^b_N - \tau^0_N e^b_M )
-   \lambda^0_b ( \tau^1_M e^b_N - \tau^1_N e^b_M )\,.
\end{array}
\end{equation}
Here, the replacement $\lambda^1_a\rightarrow \tilde{c}\, \lambda^1_a$ follows from dimensional analysis.

\subsection{Galilean longitudinal limit}
\label{sec:Gal_long_limit}

We now consider the limit where the longitudinal speed of light $\tilde{c}$ is sent to infinity.
Looking at the TSNC NG action \eqref{eq:TSNCstringwithc} above, we note the leading factor of $\tilde{c}^2$ in the integrand.
This suggests we should keep $m_{MN}=\tilde{c}^2 \tilde{m}_{MN}$ fixed in the limit.
Moreover, we should keep fixed the quantity $\tilde{c}\,\mathcal{T}$, which means the tension $\mathcal{T}$ goes to zero in the limit. 
Thus, taking the $\tilde{c}\rightarrow \infty$ limit we obtain the following Nambu--Goto-type action,
\begin{equation}
\label{GCA_NG}
S = - \frac{T_{\rm g}}{2} \int d^2 \sigma \Big[ \sqrt{-\tau}\, \tau_1^\alpha \tau_1^\beta h_{\alpha\beta} + \epsilon^{\alpha\beta} m_{\alpha\beta} \Big] \,,
\end{equation}
where we removed the tilde on $m_{MN}$ and defined the rescaled tension
\begin{equation}
\label{tension_g}
T_{\rm g} = \tilde{c}\, \mathcal{T}\,.
\end{equation}
This action \eqref{GCA_NG} can be interpreted as describing a string for which all the fluctuations have Galilean worldsheet dynamics.
Likewise, the pullbacks of the longitudinal vielbeine now induce a two-dimensional Galilean structure on the worldsheet, given by the clock one-form $\tau_\alpha^0$ and the spatial vector $\tau^\alpha_1$.
Therefore, we call the above limit a Galilean longitudinal limit.
Since we took this limit on the NG action after the speed of light limit $c\rightarrow \infty$, the above action is valid in the regime $v_{\rm char} \ll \tilde{c} \ll c$ where $v_{\rm char}$ is the characteristic velocity scale for our theory.
As we will show in~Section \ref{sec:limits_Pol}, we can also take a corresponding limit in the TSNC Polyakov action.

The symmetries of the action \eqref{GCA_NG} consist of two categories.
First, there are symmetries that originate from the large $\tilde c$ limit of the TSNC NG action \eqref{NG_TSNC}.
However, there is also an additional symmetry that arises in the limit but that cannot be viewed as originating from \eqref{NG_TSNC}.
We start by listing symmetries of the first category.
(See Appendix~\ref{app:SymAlg} for the underlying symmetry algebra of the target space geometry.)
These are the same as those of the TSNC NG action \eqref{NG_TSNC} listed in the end of Section~\ref{sec:TSNC_review}, apart from the last two, which now are replaced by
\begin{itemize}
\item Longitudinal Galilean local boost symmetry
\begin{equation}
\label{lGlb} 
\delta\tau^0_M =0 \spa 
\delta\tau^1_M= \lambda \tau^0_M\,.
\end{equation}
\item
Transverse Galilean local boost symmetry
\begin{eqnarray}
\label{stringGal_GCA}
&&\delta h_{MN} =  \lambda^0_{b} ( \tau^0_M e^b_N + \tau^0_N e^b_M )-\lambda^1_{b} ( \tau^1_M e^b_N + \tau^1_N e^b_M )\,,\nonumber\\
&&\delta m_{MN} = \lambda^1_{a} ( \tau^0_M e^a_N - \tau^0_N e^a_M )\,.
\end{eqnarray}
\end{itemize}
The new symmetry that arises in the limit is the following transformation
\begin{itemize}
\item Anisotropic Weyl transformations
\begin{equation}
\delta\tau^0_M=2\tilde\Omega\tau^0_M\,,\qquad\delta\tau^1_M=\tilde\Omega\tau^1_M\,,\qquad\delta h_{MN}=-\tilde\Omega h_{MN}\,,
\end{equation}
with $\tilde\Omega(x^M)$ an arbitrary function on the target space. This is an anisotropic Weyl transformations of $\tau^0_M$ and $\tau^1_M$ with critical exponent $z=2$. 
\end{itemize}

The above symmetries are local symmetries of the action \eqref{GCA_NG}. One finds all the possible global symmetries $\delta X^M=-\xi^M$ of the action \eqref{GCA_NG} by determining the $\xi^M$ that solves
\begin{eqnarray}
0=\delta \tau^0_M & := & \mathcal{L}_\xi\tau^0_M+\left(\Omega+2\tilde\Omega\right)\tau_M^0\,,\\ 
0=\delta \tau^1_M & := & \mathcal{L}_\xi\tau^1_M+\left(\Omega+\tilde\Omega\right)\tau_M^1+\lambda\tau^0_M\,,\\ 
0=\delta h_{MN} & := & \mathcal{L}_\xi h_{MN}-\tilde\Omega h_{MN} +\lambda^0_{b} ( \tau^0_M e^b_N + \tau^0_N e^b_M )-\lambda^1_{b} ( \tau^1_M e^b_N + \tau^1_N e^b_M )\,,\\    
0=\delta m_{MN} & := & \mathcal{L}_\xi m_{MN}+\lambda^1_{a} ( \tau^0_M e^a_N - \tau^0_N e^a_M )+\partial_M \Sigma_N - \partial_N \Sigma_M\,.
\end{eqnarray}
Below in Section \ref{sec:symmetries} we shall investigate  the global symmetries for a particular family of target space geometries.
In particular, we will see that the anisotropic Weyl symmetries will play an important role.

The above target space geometry generalizes the previously found $U(1)$-Galilean geometry discovered and studied in Refs.~\cite{Harmark:2017rpg,Harmark:2018cdl,Harmark:2019upf,Harmark:2020vll}.
The $U(1)$-Galilean geometry corresponds to singling out a particular target space direction $v$, so that
\begin{equation}
    x^M = (x^\mu, v)\,,
\end{equation}
with $\mu = 0,1,...,d$. In this identification, the $v$ direction is required to be an isometry. Furthermore, one requires
\begin{equation}
    \tau^0_M = (\tau^0_\mu ,0 ) \spa \tau^1_M = (\tau^1_\mu , 1) \spa e^a_M = (e^a_\mu ,0 )\,,
\end{equation}
which means that $v$ is a longitudinal direction. 
The $U(1)$ in the $U(1)$-Galilean geometry refers to the $U(1)$ one-form gauge field $m_\mu = - m_{v\mu}$ which is the part of the $m_{MN}$ two-form that has one leg along the $v$-direction.

As explained in Refs.~\cite{Harmark:2017rpg,Harmark:2018cdl,Harmark:2019upf,Harmark:2020vll}, the $U(1)$-Galilean geometry naturally arises in the context of the Spin Matrix theory limit of the AdS/CFT correspondence \cite{Harmark:2014mpa}, in which one finds sigma models of the type \eqref{GCA_NG} from limits of both sides of the correspondence.
Thus, the target space geometry presented in this section is a generalization of the target space geometry found from the Spin Matrix theory limit of the AdS/CFT correspondence.
As we will see in Section~\ref{sec:SMT_target}, the $U(1)$ gauge field $m_\mu$ then plays a crucial role in the dynamics of the string.
For more detail on the derivation of these backgrounds, including their relation to Penrose limits, see~\cite{Harmark:2020vll}.

\subsection{Carrollian longitudinal limit}
\label{sec:Car_long_limit}

Interestingly, there is also another natural limit of the string action~\eqref{eq:TSNCstringwithc} that one can take: sending the longitudinal speed of light $\tilde{c}$ to zero. 
In this limit we keep all the fields fixed and keep the following tension fixed
\begin{equation}
\label{eq:tension_c}
    T_c = \frac{\mathcal{T}}{\tilde{c}}
\end{equation}
to obtain the action 
\begin{equation}
\label{BMS_NG}
S =  \frac{T_c}{2} \int d^2 \sigma \Big[  \sqrt{-\tau}\, \tau_0^\alpha \tau_0^\beta h_{\alpha\beta} - \epsilon^{\alpha\beta} m_{\alpha\beta} \Big] \,,
\end{equation}
from the $\tilde{c}\rightarrow 0$ limit of the TSNC action \eqref{eq:TSNCstringwithc}.
This Nambu--Goto-type action can be interpreted as describing a string with Carrollian worldsheet dynamics for the string fluctuations.
Likewise, the pullbacks of the longitudinal vielbeine now induce a two-dimensional Carrollian structure on the worldsheet, given by the time vector $\tau^\alpha_0$ and the spatial one-form $\tau_\alpha^1$.
See Appendix~\ref{app:SymAlg} for the underlying symmetry algebra of the resulting target space geometry.
This action is valid in the regime $\tilde{c} \ll v_{\rm char} \ll c$ where $v_{\rm char}$ is the characteristic velocity scale for our theory.
Again, we consider the same limit in terms of the TSNC Polyakov action in Section~\ref{sec:limits_Pol} below.

The symmetries of the action \eqref{BMS_NG} are the same as for the TSNC NG action \eqref{NG_TSNC} listed in the end of Section~\ref{sec:TSNC_review}, apart from the last two, which now are 
\begin{itemize}
\item Longitudinal Carrollian local boost symmetry
\begin{equation}
\label{lClb} 
\delta\tau^0_M= \lambda \tau^1_M \spa 
\delta\tau^1_M=0\,. 
\end{equation}
\item
Transverse Galilean local boost symmetry
\begin{eqnarray}
\label{stringGal_BMS}
\delta h_{MN} & = &  \lambda^0_{b} ( \tau^0_M e^b_N + \tau^0_N e^b_M )-\lambda^1_{b} ( \tau^1_M e^b_N + \tau^1_N e^b_M )
\,,\nonumber\\
\delta m_{MN} & = & -\lambda^0_{b} ( \tau^1_M e^b_N - \tau^1_N e^b_M )\,.
\end{eqnarray}
\end{itemize}
There is also a new local target space symmetry which is given by
\begin{itemize}
\item Anisotropic Weyl transformations
\begin{equation}
    \delta\tau^0_M=\tilde\Omega\tau^0_M\,,\qquad\delta\tau^1_M=2\tilde\Omega\tau^1_M\,,\qquad\delta h_{MN}=-\tilde\Omega h_{MN}\,,
\end{equation}
with $\tilde\Omega(x^M)$ an arbitrary function on the target space. 
\end{itemize}
A global symmetry of the theory \eqref{BMS_NG} is a target space diffeomorphism $\delta X^M=-\xi^M$ which leaves the couplings in \eqref{BMS_NG} invariant up to a local gauge transformation.
This means that for a global symmetry we require that $\xi^M$ satisfies
\begin{eqnarray}
   0= \delta\tau^0_M & := & \mathcal{L}_\xi \tau^0_M+\left(\Omega+\tilde\Omega\right)\tau^0_M+\lambda\tau^1_M\,,\label{eq:sym1a}\\
    0=\delta\tau^1_M & := & \mathcal{L}_\xi \tau^1_M+\left(\Omega+2\tilde\Omega\right)\tau^1_M\,,\\
   0= \delta h_{MN} & := & \mathcal{L}_\xi h_{MN}-\tilde\Omega h_{MN}+\lambda^0_{b} ( \tau^0_M e^b_N + \tau^0_N e^b_M )-\lambda^1_{b} ( \tau^1_M e^b_N + \tau^1_N e^b_M )\,,\\
   0= \delta m_{MN} & := & \mathcal{L}_\xi m_{MN}-\lambda^0_{b} ( \tau^1_M e^b_N - \tau^1_N e^b_M )+\partial_M\Sigma_N-\partial_N\Sigma_M\,.\label{eq:sym4}
\end{eqnarray}

The Nambu--Goto actions for the Galilean \eqref{GCA_NG} and the Carrollian limit \eqref{BMS_NG} are related by the map
\begin{equation}
\label{relation_NG}
 \tau^0_M \leftrightarrow \tau^1_M \,,
\end{equation}
while keeping $h_{MN}$,  and $m_{MN}$ fixed.
We thus see that these two theories are (locally) related by an interchange of worldsheet time and space.
This is closely related  to the known duality between the notions of Newton-Cartan and Carrollian geometry in two dimensions.

\subsection{Longitudinal limits of the non-relativistic Polyakov action}
\label{sec:limits_Pol}

Above we have presented the longitudinal Galilean and Carrollian limits of the TSNC NG action \eqref{NG_TSNC}. We now present the equivalent limits in the Polyakov formulation. The TSNC Polyakov action (again without the dilaton term), which follows from an infinite speed of light limit $c\rightarrow \infty$ of the relativistic Polyakov action, is \cite{Bidussi:2021ujm}
\begin{equation}
\label{Pol_TSNC}
S
= - \frac{\mathcal{T}}{2} \int d^2 \sigma  \Big[  \rho\, \eta^{AB} \rho_A^\alpha \rho_B^\beta h_{\alpha\beta} +\epsilon^{\alpha\beta} m_{\alpha\beta}
+ \lambda \epsilon^{\alpha\beta}  \rho^+_\alpha \tau^+_\beta + \bar\lambda\epsilon^{\alpha\beta}  \rho^-_\alpha \tau^-_\beta \Big]\,,
\end{equation}
where $\mathcal{T} = c \, T_{\rm F}$ is fixed for $c\rightarrow \infty$.
Here the worldsheet metric $\gamma_{\alpha\beta}$ is written in terms of worldsheet zweibeine $\rho^A_\alpha$ such that
$\gamma_{\alpha\beta}=\rho^A_\alpha \rho^B_\beta \eta_{AB}$ and we defined 
\begin{equation}
\label{zweibeine_defs}
    \rho=\epsilon^{\alpha\beta} \rho^0_\alpha \rho^1_\beta\spa \rho_0^\alpha = \frac{1}{\rho}\epsilon^{\alpha\beta} \rho^1_\beta \spa \rho_1^\alpha =- \frac{1}{\rho}\epsilon^{\alpha\beta} \rho^0_\beta\,.
\end{equation}
as well as $\rho^\pm_\alpha = \rho^0_\alpha \pm \rho^1_\alpha$ and $\tau^{\pm}_M =\tau^0_M \pm \tau^1_M$. Moreover, $\lambda$ and $\bar\lambda$ are Lagrange multipliers.
The action \eqref{Pol_TSNC} follows from  an infinite speed of light limit $c\rightarrow \infty$ of the relativistic Polyakov action (without dilaton term), a limit which is equivalent to the one we reviewed for the NG action in Section~\ref{sec:TSNC_review} \cite{Bidussi:2021ujm}.

Making now the longitudinal speed of light $\tilde{c}$ manifest, as we did for the NG action in Section~\ref{sec:long_speed_light}, the TSNC Polyakov action (without dilaton) can be written as
\begin{align}\label{Pol_TSNC2}
S
&= - \frac{\mathcal{T}}{2\, \tilde{c}} \int d^2 \sigma  \left[ \rho\left(-\rho_0^\alpha \rho_0^\beta+\tilde c^2 \rho_1^\alpha \rho_1^\beta\right) h_{\alpha\beta} +\epsilon^{\alpha\beta}m_{\alpha\beta}\right.
\\ \nonumber
&\left.\qquad\qquad\qquad
+ \omega\epsilon^{\alpha\beta}\left( \tilde{c}^2 \rho^0_\alpha \tau^0_\beta +  \rho^1_\alpha \tau^1_\beta\right)+ \psi\epsilon^{\alpha\beta} \tilde{c}^2\left(\rho^0_\alpha\tau^1_\beta+\rho^1_\alpha\tau^0_\beta\right)  \right]\,,
\end{align}
where we made the replacement $\rho^1_M \rightarrow \tilde{c}^{-1} \rho^1_M$ in line with \eqref{new_tau1} and we redefined the  Lagrange multipliers as follows
\begin{equation}
\omega = \frac{1}{\tilde{c}}( \lambda+ \bar{\lambda}) \spa \psi =\frac{1}{\tilde c^2}\left( \lambda - \bar{\lambda}\right)\,.
\end{equation}

We now take the Galilean longitudinal limit of \eqref{Pol_TSNC2} by sending $\tilde{c}\rightarrow \infty$. As in Section~\ref{sec:Gal_long_limit} we keep $\tilde{m}_{MN} = \tilde{c}^2 m_{MN}$ fixed in the limit, and subsequently remove the tilde. All other fields are kept fixed in the limit, and we keep fixed the rescaled tension $T_{\rm g}$ defined in \eqref{tension_g} as in Section~\ref{sec:Gal_long_limit}. Doing so we obtain
\begin{equation}
\label{GCA_Pol}
S = - \frac{T_{\rm g}}{2} \int d^2\sigma \Big[ \left( \rho\, \rho_1^\alpha \rho_1^\beta h_{\alpha\beta} + \epsilon^{\alpha\beta} m_{\alpha\beta} \right)
 + \omega \epsilon^{\alpha\beta} \rho^0_\alpha \tau^0_\beta + \psi \epsilon^{\alpha \beta} ( \rho^0_\alpha \tau^1_\beta + \rho^1_\alpha \tau^0_\beta) \Big]\,.
\end{equation}
The action \eqref{GCA_Pol} has a Galilean worldsheet structure, in that it is invariant under local Galilean boost transformations with parameter $\hat f$, as well as local Weyl transformations with parameter $f$, acting on the zweibeine as
\begin{equation}
\label{weyl_galilean_transf}
    \rho^0_\alpha \rightarrow f \rho^0_\alpha \spa \rho^1_\alpha \rightarrow f \rho^1_\alpha + \hat{f} \rho^0_\alpha \spa \omega\rightarrow \frac{1}{f} \omega - \frac{\hat{f}}{f^2} \psi
\spa \psi \rightarrow \frac{1}{f}\psi\,.
\end{equation}

Analyzing the action \eqref{GCA_Pol} further, we see that the Lagrange multipliers $\omega$ and $\psi$ give the constraints
\begin{equation}
\label{GCA_constraints}
\epsilon^{\alpha\beta} \rho^0_\alpha \tau^0_\beta = 0 \spa \epsilon^{\alpha \beta} ( \rho^0_\alpha \tau^1_\beta + \rho^1_\alpha \tau^0_\beta) = 0\,.
\end{equation}
These equations relate the worldsheet vielbeine to the longitudinal target space vielbeine.
Their general solution is $\rho^0_\alpha = h \tau^0_\alpha$ and $\rho^1_\alpha=h\tau^1_\alpha+\hat{h}\tau^0_\alpha$ for arbitrary functions $h$ and $\hat{h}$ on the worldsheet, which determine the worldsheet vielbeine up to the Galilean boosts and Weyl transformations.
If one substitutes this solution in the Polyakov action~\eqref{GCA_Pol} one regains the NG action \eqref{GCA_NG}.

The equations of motion from varying the zweibeine $\rho^0_\alpha$ and $\rho^1_\alpha$ are
\begin{equation}
\label{eq_e0}
 (2  \delta_\alpha^\beta    -   \rho^1_{\alpha}    \rho_1^\beta    ) \rho_1^{\gamma} h_{\beta\gamma}  + \omega \tau^0_\alpha +\psi \tau^1_\alpha = 0
\,,
\end{equation}
\begin{equation}
\label{eq_e1}
\rho^0_\alpha \rho_1^\beta \rho_1^\gamma 
h_{\beta\gamma} 
+ \psi \tau^0_\alpha = 0\,.
\end{equation}
There are four independent equations contained in \eqref{eq_e0} and \eqref{eq_e1}.
Two of them can be seen to fix the Lagrange multipliers. Contracting \eqref{eq_e0} and \eqref{eq_e1} with $\tau_0^\alpha$ we find
\begin{eqnarray}
\omega & = & - \tau_0^\alpha  (2  \delta_\alpha^\beta    -   \rho^1_{\alpha}    \rho_1^\beta    )\rho_1^{\gamma}h_{\beta\gamma}\,,\label{omega_value}\\
\psi & = & -\tau_0^\alpha \rho^0_\alpha \rho_1^\beta \rho_1^\gamma 
h_{\beta\gamma}\,.\label{psi_value}
\end{eqnarray}
The two remaining independent equations of \eqref{eq_e0} and \eqref{eq_e1} give the constraints \eqref{GCA_constraints}. To see this, we note that contracting \eqref{eq_e1} with $\rho_1^\alpha$ gives $\psi \rho_1^\alpha \tau^0_\alpha = 0$ with is equivalent to the first constraint. The second constraint is instead obtained by contracting  \eqref{eq_e0} with $\rho_1^\alpha$ and subtracting from this the contraction of \eqref{eq_e1} with $\rho_0^\alpha$.

Regarding the symmetries of the action \eqref{GCA_Pol}, one has 
worldsheet diffeomorphisms \eqref{SYM_WSdiff}, one-form gauge transformations of the two-form \eqref{SYM_twoform}, Weyl transformations of the longitudinal vielbeine \eqref{SYM_Weyl} and 
transverse Galilean local boost symmetry \eqref{stringGal_GCA}.
Furthermore, one has the symmetries
\begin{itemize}
    \item Longitudinal Galilean local boost symmetry
    \begin{equation}
\label{lGlb2} 
\delta\tau^0_M =0 \spa 
\delta\tau^1_M= \lambda \tau^0_M\,,
\end{equation}
where the other fields remain the same and $\lambda$ is an arbitrary function on the target space.
Following the constraints~\eqref{GCA_constraints},
one needs in addition
\begin{equation}
  \delta\rho^0_\alpha =0 \spa 
\delta\rho^1_\alpha= (\lambda \circ X )\rho^0_\alpha\,,
\end{equation}
where $\lambda \circ X$ is the pullback of the boost parameter $\lambda$ to the worldsheet.
\item Anisotropic Weyl transformations with critical exponent $z=2$
\begin{equation}
\delta\tau^0_M=2\tilde\Omega\tau^0_M\,,\qquad\delta\tau^1_M=\tilde\Omega\tau^1_M\,,\qquad\delta h_{MN}=-\tilde\Omega h_{MN}\,,
\end{equation}
with $\tilde\Omega(x^M)$ an arbitrary function on the target space. In addition, one needs
\begin{equation}
\delta\rho^0_\alpha =2(\tilde\Omega \circ X)\rho^0_\alpha \,,\qquad\delta\rho^1_\alpha =(\tilde\Omega \circ X)\rho^1_\alpha \,.
\end{equation}
\end{itemize}
\noindent
These symmetries are accompanied by corresponding transformation of the Lagrange multipliers (see e.g. \cite{Harmark:2018cdl,Harmark:2019upf}), but since they will not play a role in this paper we omit their details. 

For the Carrollian longitudinal limit of the TSNC Polyakov action \eqref{Pol_TSNC2} we take the limit $\tilde{c}\rightarrow 0$ while keeping the fields as well as the tension $T_c$ from~\eqref{eq:tension_c} fixed. An exception is the Lagrange multiplier $\psi$ for which we keep fixed $\tilde{\psi}=\tilde{c}^2 \psi$ and remove the tilde after the limit. 
We get
\begin{equation}
\label{BMS_Pol}
S = - \frac{T_c}{2} \int d^2\sigma \Big[ \left( -\rho\, \rho_0^\alpha \rho_0^\beta h_{\alpha\beta} + \epsilon^{\alpha\beta} m_{\alpha\beta} \right)
 + \omega \epsilon^{\alpha\beta} \rho^1_\alpha \tau^1_\beta + \psi \epsilon^{\alpha \beta} ( \rho^0_\alpha \tau^1_\beta + \rho^1_\alpha \tau^0_\beta) \Big]\,,
\end{equation}
where we dropped the tildes.
This action has a Carrollian worldsheet structure which is related to the target space longitudinal Carrollian structure by the constraints.
It is invariant under local Carrollian boost transformations with parameter $\hat f$, as well as Weyl transformations with parameter $f$, acting on the zweibeine as
\begin{equation}
\label{weyl_carroll_transf}
    \rho^0_\alpha \rightarrow f \rho^0_\alpha + \hat{f} \rho^1_\alpha \spa \rho^1_\alpha \rightarrow f \rho^1_\alpha \spa   \omega\rightarrow \frac{1}{f} \omega - \frac{\hat{f}}{f^2} \psi
\spa \psi \rightarrow \frac{1}{f}\psi\,.
\end{equation}
It is straightforward to find the symmetries of the action \eqref{BMS_Pol} just as we did for the Galilean counterpart \eqref{GCA_Pol} above.

To relate the Galilean and Carrollian strings, one exchanges both the longitudinal target space vielbeine as in \eqref{relation_NG} as well as the worldsheet vielbeine,
\begin{equation}
\label{relation_Pol}
\rho^0 \leftrightarrow \rho^1 \,,
\end{equation}
while keeping $h_{MN}$ and $m_{MN}$ fixed.
In this way, the two Lagrangians \eqref{GCA_Pol} and \eqref{BMS_Pol} map onto each other. This is in accordance with the expected map between Galilean and Carrollian worldsheet theories.
With this, one can also map the analysis of the constraints and Lagrange multipliers in \eqref{eq_e0}, \eqref{eq_e1}, \eqref{omega_value} and \eqref{psi_value} for the Galilean string to the case of the Carrollian string, as well as the local symmetries of the action.

\section{Worldsheet gauge fixings}
\label{sec:ws_gauge_fixings}

In this section we describe two different gauge fixings of the worldsheet diffeomorphisms. 
In Section \ref{sec:static_gauge} we present a full gauge fixing of both the Galilean and Carrollian Nambu--Goto actions introduced in Section \ref{sec:limits}.
Additionally, in Section \ref{sec:galilean_gauge} we present a particular partial gauge fixing of Galilean Nambu--Goto and Polyakov actions.
We shall employ both of these gauge fixings in Section \ref{sec:SMT_target} for a specific family of target space geometries.

\subsection{Nambu--Goto actions: static gauge}
\label{sec:static_gauge}

In this section we identify a natural gauge choice for the Nambu--Goto actions introduced in Section \ref{sec:limits} which fixes (almost) all of the worldsheet diffeomorphisms and leaves only manifestly physical degrees of freedom.
We show that one can always go to this gauge, which is a version of the static gauge adapted to our situation where strings necessarily wind along a particular direction.
Note that, in contrast to the relativistic Nambu--Goto action,
the usual square root in the action is no longer present when this gauge is applied
after the $c\rightarrow \infty$ limit.
This holds for all three non-relativistic NG actions that we discussed in Section \ref{sec:limits}, i.e. the TSNC action~\eqref{NG_TSNC}, the Galilean action~\eqref{GCA_NG} and the Carrollian action~\eqref{BMS_NG}.
As a result, this static gauge is a natural choice for the NG actions and is a good starting point for a quantization procedure.

To begin, 
we introduce the following notation for the target space coordinates. We will denote the coordinates as $x^M = (t,v,x^i)$ where $t$ is the time coordinate, $v$~parametrizes a compact direction  with radius $R_v$ that the string wraps, and where the remaining coordinates are denoted by $x^i$, with $i=1,...,d$. The associated embedding scalars will be denoted by $X^t, X^v, X^i$ where $X^t$ and $X^i$ are periodic functions of~$\sigma$ and where $X^v\sim X^v+2\pi w R_v$ with $w$ the winding number.
We shall use this notation in this section as well as in Section \ref{sec:SMT_target}.

We wish to gauge fix the worldsheet diffeomorphisms
\begin{equation}
  (\tau,\sigma)\mapsto (\tau'(\tau,\sigma),\sigma'(\tau,\sigma))\,,
\end{equation}
where $-\infty<\tau,\tau'<\infty$ and where we take $\sigma,\sigma'\in [0,2\pi)$, keeping the origin and the periodicity of the latter fixed.
Infinitesimally, we have generators $\delta\sigma^\alpha=-\zeta^\alpha$ where the $\zeta^\alpha$ are periodic in $\sigma$ and where $\zeta^\sigma$ cannot be a function of $\tau$ only since this would shift the origin of the $\sigma$ coordinate.

One way to achieve this is to choose a gauge that is similar to static gauge for the relativistic string,%
\footnote{%
  Since we require the string to lie along the $(t,v)$ directions, we must have $\pd_\tau X^t \pd_\sigma X^v - \pd_\sigma X^t \pd_\tau X^v\neq 0$ so that the embedding is non-singular,
  which is respected by the gauge choice~\eqref{eq:staticgauge} above.
}
\begin{equation}\label{eq:staticgauge}
    X^t=K\tau\,,\qquad X^v=wR_v\sigma+f(\tau)\,,
\end{equation}
where $w\neq 0$ is the integer winding number of the string along the compact $v$~direction which is a circle of radius $R_v$, and where furthermore $f$ is an arbitrary function of $\tau$. The constant $K>0$
is there on dimensional grounds since $\tau$ is dimensionless and $X^t$~has dimensions of time.
We will refer to \eqref{eq:staticgauge} as the {\sl static gauge}.

We need to show that we can always reach this gauge, or, equivalently, that starting from this gauge choice one can generate any $X^t$ and $X^v$. We thus want to show that starting from the particular embedding~\eqref{eq:staticgauge}
we can reach the most general (infinitesimal) embedding\footnote{The gauge \eqref{eq:generalsol} reflects the requirements that $X^t$ is a periodic function of $\sigma$ and that $X^v\sim X^v+2\pi wR_v$ when $\sigma\sim\sigma+2\pi$. }
\begin{equation}\label{eq:generalsol}
    \bar X^t = G(\tau)+G_{\text{per}}(\tau,\sigma)\,,\qquad \bar X^v=wR_v\sigma+\tilde f(\tau)+F_{\text{per}}(\tau,\sigma)\,,
\end{equation}
where the two are related by a worldsheet diffeomorphism 
\begin{equation}\label{eq:reachinggc}
     X^t+\delta X^t=\bar X^t\,,\qquad  X^v+\delta X^v=\bar X^v\,.
\end{equation}
In here 
\begin{equation}
    \delta  X^t=\zeta^\alpha\partial_\alpha  X^t\,,\qquad\delta  X^v=\zeta^\alpha\partial_\alpha X^v\,.
\end{equation}
Furthermore, the functions $F_{\text{per}}$ and $G_{\text{per}}$ are infinitesimal arbitrary periodic functions of $(\sigma,\tau)$ of the form
\begin{equation}\label{eq:periodic}
    \sum_{n\neq 0} a_n(\tau) e^{in\sigma}\,,
\end{equation}
where $a_{-n}=a_n^\star$ and where the sum is over all nonzero $n\in\mathbb{Z}$. 
The conditions \eqref{eq:reachinggc} amount to 
\begin{equation}
    K\tau+K\zeta^\tau=G(\tau)+G_{\text{per}}(\tau,\sigma)\,,\qquad f+\zeta^\tau \pd_\tau f+wR_v\zeta^\sigma=\tilde f+F_{\text{per}}(\tau,\sigma)\,.
\end{equation}
We can solve these equations for $\zeta^\tau$, $\tilde f$ and $\zeta^\sigma$ with the latter being of the form \eqref{eq:periodic}. 
(We can solve for $\tilde f$ in terms of $f$ because $f$ is already arbitrary.)
This shows one can reach any other gauge starting from the static gauge \eqref{eq:staticgauge}.

One can also show that the gauge fixings of $X^t$ and $X^v$ in the static gauge \eqref{eq:staticgauge} do not give rise to additional equations of motion, other than the ones of the gauge fixed action, in contrast to for example the conformal gauge in the usual relativistic string.
To see this we observe that the Lagrangians studied in this paper are of the form $\mathcal{L}(X^M,\partial_\alpha X^M)$. If we perform a gauge transformation of the form $\delta X^M=\zeta^\alpha\partial_\alpha X^M$, i.e. a 2-dimensional diffeomorphism generated by $\zeta^\alpha$, then the Lagrangian will transform as a density, i.e. as $\delta\mathcal{L}=\partial_\alpha\left(\zeta^\alpha\mathcal{L}\right)$. Using standard manipulations this leads to
\begin{equation}\label{eq:2ddiffeos}
    E_M\zeta^\beta\partial_\beta X^M=\partial_\alpha\left( T^\alpha{}_\beta \zeta^\beta\right)\,,
\end{equation}
where we defined
\begin{eqnarray}
    E_M & = & \partial_\alpha\left(\frac{\partial\mathcal{L}}{\partial \partial_\alpha X^M}\right)-\frac{\partial\mathcal{L}}{\partial X^M}\,,\\
    T^\alpha{}_\beta & = & \frac{\partial\mathcal{L}}{\partial \partial_\alpha X^M}\partial_\beta X^M-\delta^\alpha_\beta\mathcal{L}\,.
\end{eqnarray}
The equations of motion of the embedding scalars $X^M$ are given by $E_M=0$ and $T^\alpha{}_\beta$ is a worldsheet energy-momentum tensor. Gauge invariance under 2-dimensional diffeomorphisms requires that \eqref{eq:2ddiffeos} is satisfied for any $\zeta^\alpha$, which implies%
\footnote{%
  This follows from first looking at constant but otherwise arbitrary $\zeta^\alpha$.
  Substituting the resulting condition in \eqref{eq:2ddiffeos} tells us that we need \eqref{eq:T=0}.
}
\begin{equation}\label{eq:T=0}
    T^\alpha{}_\beta=0\,.
\end{equation}
Equation \eqref{eq:2ddiffeos} then simplifies to
\begin{eqnarray}
    0 & = & E_t\dot X^t+E_v\dot X^v+E_i\dot X^i\,,\\
    0 & = & E_t X'^t+E_v X'^v+E_i X'^i\,,
\end{eqnarray}
which expresses the well-known fact that the equations of motion are not all independent (due to the gauge symmetry).
Here, we write $\pd_\tau X^M = \dot{X}^M$ and $\pd_\sigma X^M = X'^i$ as usual.
We thus see that in static gauge the equations of motion of $X^t$ and $X^v$ are automatically solved since we can solve the above two equations for $E_t$ and $E_v$.
Using any of the Nambu--Goto Lagrangians in this paper one can explicitly verify that equation \eqref{eq:T=0} is identically solved. 

Again, while static gauge is equally valid for the usual relativistic string theory, the fact that the non-relativistic Nambu--Goto actions discussed above do not have a square root after this gauge fixing means that this gauge is particularly appealing in the present setting, as we will see below.

\subsection{Galilean string: partial gauge fixing}
\label{sec:galilean_gauge}

In this section we perform a partial gauge fixing of the Nambu--Goto and Polyakov actions of the Galilean string.
The goal of this gauge fixing is to get rid of certain negative definite kinetic terms in the action.
Without these terms the quantization can proceed significantly more efficiently.
Note that one cannot in general employ this partial gauge fixing at the same time as the static gauge of Section \ref{sec:static_gauge}.
However, it will be possible to do so for the families of target space geometries we consider in Section~\ref{sec:SMT_target}.

We consider the Nambu--Goto action~\eqref{GCA_NG}
of the Galilean string theory whose Lagrangian we repeat here for convenience,
\begin{equation}
\label{L_GCA_NG}
\CL =   - \frac{T_g}{2}  \sqrt{-\tau}\, \tau_1^\alpha \tau_1^\beta h_{MN} \partial_\alpha X^M \partial_\beta X^N - T_g \, m_{MN} \dot{X}^M {X'}^N \,,
\end{equation} 
where 
\begin{equation}
\tau_1^\alpha = - \frac{1}{\sqrt{-\tau}} \epsilon^{\alpha\beta} \tau^0_\beta\,,\qquad\sqrt{-\tau}=\epsilon^{\alpha\beta}\tau^0_\alpha\tau^1_\beta=\tau^0_\tau\tau^1_\sigma-\tau^0_\sigma\tau^1_\tau\,.
\end{equation}
Explicitly, the Lagrangian \eqref{L_GCA_NG} can be rewritten as 
\begin{equation}
\CL =   - \frac{T_g}{2}\frac{1}{\sqrt{-\tau}}h_{MN}\left(\dot X^M\tau^0_\sigma-X'^M\tau^0_\tau\right)\left(\dot X^N\tau^0_\sigma-X'^N\tau^0_\tau\right)  - T_g \, m_{MN} \dot{X}^M {X'}^N \,.
\end{equation} 
Now let us look at the structure of the kinetic part of this Lagrangian, {\sl i.e.}~terms that include $\dot{X}^M$. 
Note that this action only has terms quadratic in $\dot{X}^M$ provided that $\tau^0_\sigma\neq 0$, which means that the pullback of the target space longitudinal one-form $\tau^0_M dx^M$ to the worldsheet has nonzero $d\sigma$ component.
However, the resulting kinetic term has an opposite sign, which would seemingly lead to negative-norm states.
Therefore, we are led to considering worldsheet coordinates such that $\tau^0_M X'^M=0$, or in other words
\begin{equation}
  \label{GCA_con}
  \tau^0_M {X'}^M =0
  \quad\Longleftrightarrow\quad
  \tau^0_\alpha d\sigma^\alpha
  = N(\tau,\sigma) d\tau\,.
\end{equation}
From the perspective of the Galilean structure on the worldsheet, this condition is very natural: it requires that the coordinates $(\tau,\sigma)$ are adapted to the one-dimensional spatial foliation of equal-$\tau$ surfaces that $\tau^0$ defines.
It is then not surprising that such adapted coordinates significantly simplify the description of the dynamics of the worldsheet.

We impose this condition using a constraint with a Lagrange multiplier $\hat\omega$, so that
the Nambu--Goto Lagrangian \eqref{L_GCA_NG} for the Galilean string gets modified to
\begin{equation}
\label{L_GCA_NG2}
\CL =   - \frac{T_g}{2} \left[ \frac{\tau^0_P \dot{X}^P}{\tau^1_Q {X'}^Q} h_{MN} {X'}^M  {X'}^N +2 m_{MN} \dot{X}^M {X'}^N + \hat{\omega}\, \tau^0_M {X'}^M \right]\,.
\end{equation} 
Note that by imposing the condition~\eqref{GCA_con} we are restricting part of the diffeomorphism symmetry on the worldsheet.
However, it can be shown that the coordinate choice~\eqref{GCA_con} greatly simplifies the subsequent Hamiltonian analysis.
Since this theory is now first-order in time derivatives, it turns out that part of $m_{MN}$ plays the role of a symplectic potential on the field space described by the embedding fields $X^M$, as we will see in detail for the particular backgrounds we consider in Section~\ref{sec:SMT_target}.

For the above reasons, we will stick to worldsheet coordinates of the type \eqref{GCA_con}, although we believe (but will not check) that a careful analysis of the general case should lead to equivalent results.

A similar result can be obtained from the Galilean Polyakov action~\eqref{GCA_Pol}, where we see that the $\omega$ Lagrange multiplier gives the constraint
\begin{equation}
e^0_\tau \tau^0_\sigma = e^0_\sigma \tau^0_\tau\,.
\end{equation}
This suggests that our previous assumption that $\tau^0_\sigma=0$
for the Nambu--Goto action
(and thus $\tau^0_\tau\neq 0$ since we assume that $\tau^0_\alpha$ is nowhere vanishing on the worldsheet)
can be achieved by setting $e^0_\sigma=0$ in the Polyakov case. 
Again, this can be interpreted as a choice of coordinates that is adapted to the Galilean structure on the worldsheet, as we discussed below~\eqref{GCA_con}.
Setting $e^0_\sigma=0$ by hand in \eqref{GCA_Pol} one finds the Lagrangian%
\footnote{Note that one needs $e^0_\tau\neq 0$ to have a well-defined worldsheet theory. Note also that $e_1^\tau =0$ and $e_1^\sigma = 1/e^1_\sigma$.}
\begin{equation}
\label{L_GCA_restr1}
\CL = - \frac{T_g}{2} \left[  \frac{e^0_\tau}{e^1_\sigma}  h_{MN}{X'}^M {X'}^N + 2 m_{MN}  \dot{X}^M {X'}^N
 + \omega  e^0_\tau \tau^0_\sigma + \psi  (e^0_\tau \tau^1_\sigma + \epsilon^{\alpha \beta} e^1_\alpha \tau^0_\beta) \right]\,.
\end{equation}
Using the redefinition
$\hat{\omega} = \omega e^0_\tau + \psi e^1 _\tau$
we then get
\begin{equation}
\label{L_GCA_restr2}
\CL = - \frac{T_g}{2} \left[  \frac{e^0_\tau}{e^1_\sigma}  h_{MN}{X'}^M {X'}^N + 2 m_{MN}  \dot{X}^M {X'}^N
 + \hat{\omega} \tau^0_\sigma + \psi  (e^0_\tau \tau^1_\sigma - e^1_\sigma \tau^0_\tau) \right]\,.
\end{equation}
Integrating out $\psi$ one finds \eqref{L_GCA_NG2}, thus one can think of \eqref{L_GCA_restr2} as a Polyakov version of the Lagrangian \eqref{L_GCA_NG2}.

One can find the following equations of motion by varying $e^0_\tau$ and $e^1_\sigma$
\begin{equation}
 \frac{1}{e^1_\sigma}  h_{MN}{X'}^M {X'}^N + \psi \tau^1_\sigma=0
\spa
 \frac{e^0_\tau}{(e^1_\sigma)^2}  h_{MN}{X'}^M {X'}^N + \psi \tau^0_\tau=0\,.
\end{equation}
Multiplying the first equation with $-e^0_\tau /e^1_\sigma$ and adding this to the second equation we find again the constraint that is obtained from varying $\psi$ in \eqref{L_GCA_NG2}. The other independent equation determines the value of $\psi$.

Comparing \eqref{L_GCA_restr2} with \eqref{BMS_Pol} we see that the difference is that the former has one worldsheet zweibein component less and so there is no equation of motion obtained by varying $e^0_\sigma$ since this field is absent. Varying  $e^0_\sigma$ in \eqref{BMS_Pol} leads to an equation that determines the Lagrange multiplier $\hat\omega$. 

For completeness, let us see explicitly what it means for the worldsheet diffeomorphisms to restrict $e^0_\sigma=0$ as we did above.
Consider a general worldsheet diffeomorphism
\begin{equation}
\tau \rightarrow \tilde{\tau} (\tau,\sigma) \spa \sigma  \rightarrow \tilde{\sigma} (\tau,\sigma)\,.
\end{equation}
This gives the transformation
\begin{equation}
e^0_\sigma = \frac{\partial \tilde{\tau}}{\partial\sigma} \tilde{e}^0_{\tilde{\tau}} + \frac{\partial \tilde{\sigma}}{\partial\sigma} \tilde{e}^0_{\tilde{\sigma}}\,. 
\end{equation} 
Thus, if we impose $e^0_\sigma=0$ and subsequently require that any transformation should retain $\tilde{e}^0_{\tilde{\sigma}}=0$ we get the condition
\begin{equation}
\frac{\partial \tilde{\tau}}{\partial\sigma} = 0\,.
\end{equation}
This means that the most general worldsheet diffeomorphisms respecting $e^0_\sigma=0$ are given by the foliation-preserving diffeomorphisms associated to the equal-$\tau$ spatial surfaces,
\begin{equation}
\label{diffeo_constr}
\tau \rightarrow  \tilde{\tau} (\tau) \spa \sigma  \rightarrow  \tilde{\sigma} (\tau,\sigma) \,.
\end{equation}
These are the residual gauge transformation of the gauge choice $e^0_\sigma=0$.

\section{Spin Matrix target spacetimes for the Galilean string}
\label{sec:SMT_target}

Above we have introduced new classes of non-relativistic string sigma models, resulting from both the Galilean and Carrollian longitudinal limits of the TSNC string.
In this section we focus further on the Galilean case, introducing a solvable model, corresponding to a simple class of target space geometries and sigma models that arise in the context of particular limits of AdS/CFT.
The Carrollian case is left for future work, since it requires a deeper study of what target space geometries are permitted.

\subsection{Solvable model}
As in Section \ref{sec:static_gauge},
we will use the target space coordinates $x^M = (t,v,x^i)$, where the $v$~direction is periodic with radius $R_v$ and the string winds around it $w$ times.
As a first simplifying assumption, we choose
\begin{equation}\label{eq:BMS_restrbackgr0}
    \tau^0_M=\delta^t_M \spa \tau^1_M=\delta^v_M\,.
\end{equation}
We will also assume that both $\partial_t$ and $\partial_v$ are manifest isometries so that the target space tensors $h_{MN}$ and $m_{MN}$ are functions of the $x^i$ coordinates only. 

We now impose the partial worldsheet gauge fixing \eqref{GCA_con} for the Nambu--Goto action of the Galilean string.
For a background with \eqref{eq:BMS_restrbackgr0} this amounts to the condition
\begin{equation}
\label{GCA_con_restr}
    X'{}^t = 0\,.
\end{equation}
The partially gauge fixed
NG Lagrangian \eqref{L_GCA_NG2} then reduces to
\begin{equation}
\CL =   - \frac{T_g}{2} \left[ \frac{\dot X^t}{X'^v} h_{MN} {X'}^M  {X'}^N +2 m_{MN} \dot{X}^M {X'}^N + \hat{\omega}\,X'{}^t \right]\,.
\end{equation} 
More explicitly, this becomes
\begin{eqnarray}
\label{restr_Lagr}
\CL & = &   - \frac{T_g}{2} \left[  \frac{\dot X^t}{X'^v}h_{ij} {X'}^i  {X'}^j +2 m_{iv}\left( \dot{X}^i {X'}^v-\dot{X}^v {X'}^i\right)\right.\nonumber\\
&&\left.\left(h_{vv}+2 m_{tv}\right) \dot{X}^t {X'}^v+\left(h_{vi}+2 m_{ti}\right) \dot{X}^t {X'}^i -2 m_{ij} \dot{X}^j {X'}^i+ \hat{\omega}\,X'{}^t \right]\,.\label{eq:Galstringsimplified}
\end{eqnarray}
where we absorbed terms proportional to $X'^t$ into $\hat\omega$.

To simplify our model even further we can employ the static gauge choice \eqref{eq:staticgauge}.
While the static gauge is not necessarily compatible with the partial gauge fixing~\eqref{GCA_con}, this is the case
for backgrounds with $\tau^0$ and $\tau^1$ as in~\eqref{eq:BMS_restrbackgr0}.
Indeed,
the static gauge is precisely adapted to the spacelike foliation defined by $\tau^0=dt$ in~\eqref{eq:BMS_restrbackgr0}.

The first line in \eqref{eq:Galstringsimplified} contains terms that we recognize from earlier work on Spin Matrix limits of strings on AdS$_5\times S^5$ \cite{Harmark:2017rpg,Harmark:2018cdl,Harmark:2019upf,Harmark:2020vll}.
We should consider target space geometries that are consistent with the general string dynamics, which presumably means they should be a leading-order solution to some currently-unknown vanishing beta functions equations. 
Such beta functions are expected to exist for the Galilean string since the Polyakov formulation of Section \ref{sec:limits_Pol} has a worldsheet Weyl symmetry that is not allowed to become anomalous, just as in the case of the relativistic string. However, these beta functions have not yet been computed and so we will henceforth restrict to the class of backgrounds that contain the backgrounds that were previously obtained from Spin Matrix limits of \cite{Harmark:2017rpg,Harmark:2018cdl,Harmark:2019upf,Harmark:2020vll}.
Such backgrounds arise from consistent limits of valid target space geometries in AdS/CFT, and it is therefore reasonable to expect that they should be valid backgrounds for the corresponding string sigma models.
For this reason, we will impose the following restrictions on the target space,
\begin{equation}
    h_{vv}+2 m_{tv}=0\,,\qquad h_{vi}+2 m_{ti}=0\,,\qquad m_{ij}=0\,.
\end{equation}

In the rest of Section \ref{sec:SMT_target} we shall study the solvable model 
that follows from 
the truncated theory discussed above by requiring that the Lagrangian is quadratic in $X^i$. This means that we take 
\begin{equation}
    h_{ij}=\delta_{ij}\,,\qquad F_{ij}=\partial_i m_j-\partial_j m_i=\text{cst}\,,
\end{equation}
where $F_{ij}$ is assumed to be invertible (and constant) and where we defined $m_{iv}=m_i$. This implies that the $i,j$ indices must take an even number of values. We can choose a gauge in which
$m_{i}=-\frac{1}{2}F_{ij}x^j$. 
Together with the static gauge, the gauge-fixed Galilean string Lagrangian~\eqref{restr_Lagr} then becomes
\begin{equation}\label{eq:toymodel}
\CL =  - u\, m_i \dot{X}^i + T_g \dot f m_i {X'}^i - \frac{1}{2} \tilde u^2 {X'}^i  {X'}^i \,,
\end{equation}
where for ease of notation we defined the constants
\begin{equation}
u=T_g w R_v\,,\qquad    \tilde u^2=\frac{T_g K}{wR_v}\,,
\end{equation}
and we remind the reader that $f(\tau)$ is the arbitrary zero mode in the static gauge~\eqref{eq:staticgauge}.

\subsection{Phase space}

The Lagrangian \eqref{eq:toymodel} takes the form of a phase space Lagrangian, so we can immediately read off what the corresponding Poisson brackets are. The one-form $m_i$ is a symplectic potential with $F_{ij}$ as the symplectic form, and the inverse of $F_{ij}$ is proportional to the Poisson bracket between the phase space variables which are the $X^i$. 

The Lagrangian can be written as
\begin{equation}
    L=\oint d\sigma\left(- u m_i \dot{X}^i - \frac{1}{2} \tilde u^2 {X'}^i  {X'}^i\right)+ T_g \dot f\oint d\sigma m_i {X'}^i\,.
\end{equation}
The momentum conjugate to $f$ is 
\begin{equation}
    \pi_f=\frac{\partial L}{\partial\dot f}=T_g\oint d\sigma m_i X'^i\,.
\end{equation}
Hence we have a primary constraint $\phi=\pi_f-T_g\oint d\sigma m_i X'^i\approx 0$. The total Hamiltonian 
is given by
\begin{equation}
    H=\oint d\sigma \frac{1}{2} \tilde u^2 {X'}^i  {X'}^i+\dot f \left(\pi_f-T_g\oint d\sigma m_i X'^i\right)\,,
\end{equation}
where $\dot f$ is a Lagrange multiplier. The Poisson brackets on the phase space are given by
\begin{eqnarray}
    \{ X^i(\tau,\sigma), X^j(\tau, \sigma')\} & = & -\frac{1}{u} \Delta^{ij}\delta(\sigma-\sigma')\,,\label{eq:Poissonbracket}\\
    \{ f(\tau), \pi_f(\tau)\} & = & 1\,,
\end{eqnarray}
where $\Delta^{ij}$ is the inverse of $F_{ij}$,
\begin{equation}
    F_{ij}\Delta^{jk}=\delta^{k}_i\,.
\end{equation}

The Poisson bracket between two phase space functionals $F$ and $G$ is given by
\begin{equation}
    \{ F\,,G\}=-\frac{1}{u}\oint d\sigma \Delta^{ij}\frac{\delta F}{\delta X^i(\sigma)}\frac{\delta G}{\delta X^j(\sigma)}+\frac{\delta F}{\delta f}\frac{\delta G}{\delta \pi_f}-\frac{\delta F}{\delta\pi_f}\frac{\delta G}{\delta f}\,.
\end{equation}
The equations of motion are given by
\begin{equation}
    \{ X^i(\sigma)\,,H\}=\dot X^i(\sigma)\,.
\end{equation}
This leads to
\begin{equation}\label{eq:EOMXi}
    -uF_{ij}\dot X^j+T_g\dot f F_{ij}X'^j+\tilde u^2 X''^i=0\,.
\end{equation}
Using 
\begin{equation}
    \{\pi_f\,,H\}=-\frac{\delta H}{\delta f}=\dot \pi_f=0\,,
\end{equation}
as well as the constraint enforced by $\dot f$ which gives 
\begin{equation}
    \pi_f=T_g\oint d\sigma\, m_i X'^i\,,
\end{equation}
we obtain the Euler-Lagrange equation of $f$ (see footnote \ref{fnote:EOMf} later on). This latter equation can also be shown to follow from \eqref{eq:EOMXi} by contracting \eqref{eq:EOMXi} with $X'^i$ and integrating over $\sigma$.

The constraint $\phi=\pi_f-T_g\oint d\sigma m_i X'^i\approx 0$ is first class. If we define the generator $P$ to be
\begin{equation}
    P=\frac{u}{T_g}\left(\pi_f-T_g\oint d\sigma m_i X'^i\right)\,,
\end{equation}
then we have
\begin{eqnarray}
    \delta_P X^i(\sigma) & = & \{X^i(\sigma)\,,P\}=X'^i(\sigma)\,,\\
    \delta_P f & = & \{f,P\}=wR_v\,.
\end{eqnarray}
Hence we see that $P$ generates a translation in $\sigma$ (which is equivalent to shifting $f$ since we need to preserve the gauge choice $X^v=wR_v\sigma+f$). 
This shift in $\sigma$ is at a constant $\tau$-slice. If we take the generator, $\zeta^\sigma(\tau)P$, then we get a gauge symmetry of the form $\delta\sigma=-\zeta^\sigma(\tau)$. This is of course precisely the gauge symmetry we were not able to gauge fix in the static gauge~\eqref{eq:staticgauge} because we decided to always have $\sigma\in [0,2\pi)$.

If we now go back to the Lagrangian \eqref{eq:Galstringsimplified} before imposing the static gauge, but with the target space chosen as for the solvable model, we see that the momentum conjugate to $X^v$ is
\begin{equation}
    P_v=\frac{\partial\mathcal{L}}{\partial\dot X^v}=T_gm_iX'^i\,.
\end{equation}
We thus see that $\pi_f$ is the zero mode of the $P_v$ momentum, $\pi_f=\oint d\sigma P_v$. In the quantum theory (due to single-valuedness of the string wave function) the momentum $\pi_f$ is quantized in units of $R_v$, i.e. we have
\begin{equation}
    \pi_f=\frac{n}{R_v}\,.
\end{equation}
This latter condition gives rise to the level matching condition in the quantum theory for a string that has a circle in the target spacetime with $n$ units of momentum. If we want the string to be pure winding along the $v$-direction then we need to set $\pi_f=0$. 

\subsection{Mode expansion}

The equations of motion~\eqref{eq:EOMXi} for $X^i$ are solved using a Fourier series expansion
\begin{equation}
    X^i=\sum_{n\in\mathbb{Z}} a^i_n(\tau) e^{in\left(\sigma+\frac{1}{wR_v}f\right)}\,,
\end{equation}
where
\begin{equation}\label{eq:EOMa}
    uF_{ij}\dot a^j_n+\tilde u^2 n^2 a_n^i=0\,,
\end{equation}
and where we have the reality condition $(a_n^i)^*=a_{-n}^i$. 
In accordance with the above expansion we note that the coordinate transformation $(\tau,\sigma)\rightarrow (\tilde\tau,\tilde\sigma)=(\tau,\sigma+\frac{1}{wR_v}f)$ is precisely what removes the $\dot f$ term from the equations of motion for $X^i$.
For $n=0$ we find that $\dot a_0^i=0$ so there is no centre of mass motion. 

Viewing $F_{ij}$ as a $2n \times 2n$ antisymmetric matrix, it is similar by a rotation to a block-diagonal matrix $\Sigma$ of the form
\begin{equation}
    \Sigma=\left(
\begin{array}{ccc}
\lambda_1\Omega & 0 & \cdots\\
0 & \lambda_2\Omega & \cdots\\
\vdots&\vdots&
\end{array}
    \right) \,,
\end{equation}
where $\pm i\lambda_I$ with $\lambda_I>0$ for $I=1,\ldots,n$ are the eigenvalues of $F$ and where $\Omega$ is the two by two matrix
\begin{equation}
    \Omega=\left(
\begin{array}{cc}
0 & 1 \\
-1 & 0
\end{array}
    \right)\,.
\end{equation}
Since the kinetic term in \eqref{eq:toymodel} is rotation invariant we can without loss of generality set $F=\Sigma$. 
The equations of motion \eqref{eq:EOMa} are then
\begin{eqnarray}
    u\lambda_I\dot a_n^{2I}+\tilde u^2 n^2 a_n^{2I-1} & = & 0\,,\\
    -u\lambda_I\dot a_n^{2I-1}+\tilde u^2 n^2 a_n^{2I} & = & 0\,,
\end{eqnarray}
which are solved by
\begin{eqnarray}
    a^{2I}_n & = & A^{2I}_n e^{-i\omega^I_n\tau}+B^{2I}_n e^{i\omega^I_n\tau}\,,\\
    a^{2I-1}_n & = & iA^{2I}_n e^{-i\omega^I_n\tau}-iB^{2I}_n e^{i\omega^I_n\tau}\,,
\end{eqnarray}
where we defined
\begin{equation}
    \omega^I_n=\frac{\tilde u^2 n^2}{u\lambda_I}\,,
\end{equation}
and where the reality conditions are $(A_n^{2I})^*=B^{2I}_{-n}$ and $(B_n^{2I})^*=A^{2I}_{-n}$. The Poisson bracket \eqref{eq:Poissonbracket} in terms of the $a^i_m$ leads to
\begin{equation}
    \{a^i_m\,,a^j_n\}=-\frac{1}{2\pi u}\Delta^{ij}\delta_{m+n,0}\,.
\end{equation}
Since we have set $F=\Sigma$ we have that $\Delta$ as a matrix is given by
\begin{equation}
    \Delta=\left(
\begin{array}{ccc}
-\lambda_1^{-1}\Omega & 0 & \cdots\\
0 & -\lambda_2^{-1}\Omega & \cdots\\
\vdots&\vdots&
\end{array}
    \right)\,.
\end{equation}
Using this, we obtain the following nonzero Poisson brackets between the $A^{2I}_n, B^{2I}_n$
\begin{equation}
    \{A^{2I}_m\,, B^{2I}_n\}=-\frac{i}{4\pi u\lambda_I}\delta_{m+n,0}\,,
\end{equation}
with all other Poisson brackets zero.

It is useful to define complex scalar field combinations
\begin{equation}
    \Phi^I=X^{2I-1}+iX^{2I}=\sum_{n\in\mathbb{Z}}2i A_n^{2I}e^{-i\omega^I_n\tau}e^{in\left(\sigma+\frac{1}{wR_v}f\right)}\,.
\end{equation}
These are solutions to the 2-dimensional Schr\"odinger equation
\begin{equation}
    i\partial_\tau\Phi^I-i\frac{1}{wR_v}\dot f\partial_\sigma\Phi^I=-\frac{\tilde u^2}{u\lambda_I}\partial_\sigma^2\Phi^I\,.
\end{equation}
This follows from rewriting \eqref{eq:EOMXi}. We thus have $n$ decoupled Schr\"odinger fields.

\subsection{Symmetries}
\label{sec:symmetries}

In Section~\ref{sec:Gal_long_limit} we derived that the global symmetries of the Galilean string for a given target space are the solutions for $\xi^M$ of the equations
\begin{eqnarray}
0=\delta \tau^0_M & := & \mathcal{L}_\xi\tau^0_M+\left(\Omega+2\tilde\Omega\right)\tau_M^0\,,\\ 
0=\delta \tau^1_M & := & \mathcal{L}_\xi\tau^1_M+\left(\Omega+\tilde\Omega\right)\tau_M^1+\lambda\tau^0_M\,,\\ 
0=\delta h_{MN} & := & \mathcal{L}_\xi h_{MN}-\tilde\Omega h_{MN} +\lambda^0_{b} ( \tau^0_M e^b_N + \tau^0_N e^b_M )-\lambda^1_{b} ( \tau^1_M e^b_N + \tau^1_N e^b_M )\,,\ \ \\  
0=\delta m_{MN} & := & \mathcal{L}_\xi m_{MN}+\lambda^1_{a} ( \tau^0_M e^a_N - \tau^0_N e^a_M )+\partial_M \Sigma_N - \partial_N \Sigma_M\,,
\end{eqnarray}
which we repeated here for convenience. We will now solve these equations for the target space
\begin{equation}
  \label{eq:backgrounds}
  \begin{gathered}
    \tau^0_Mdx^M=dt\,,\qquad\tau^1_Mdx^M=dv\,,
    \\
    h_{MN}dx^Mdx^N=dx^idx^i\,,\qquad \frac{1}{2}m_{MN}dx^M\wedge dx^N=m_{i}dx^i\wedge dv\,,
  \end{gathered}
\end{equation}
where
$m_{i}=-\frac{1}{2}F_{ij}x^j$
as above.
The solution for $\xi^M$ is given by
\begin{eqnarray}
    \xi^t & = & a+bt+\frac{c}{2}t^2\,,\\
    \xi^v & = & d+et+\frac{v}{2}(b+ct)\,,\\
    \xi^i & = & a^i(t,v)+\lambda^i{}_j x^j-\frac{1}{4}\left(b+ct\right)x^i\,,
\end{eqnarray}
where the $\lambda_{ij}$ are given by 
\begin{equation}
    \lambda_{ij}=\tilde\lambda_{ij}-\frac{v}{2}F_{ij}\left(e+\frac{c}{4}v\right)\,,
\end{equation}
and where $a^i(t,v)$ obeys
\begin{equation}\label{eq:conditionai}
    F_{ij}\partial_t a^j-\partial_v^2 a^i=0\,.
\end{equation}
The parameters $a,b,c,d,e,\tilde\lambda_{ij}$ are constants with the latter obeying
\begin{equation}\label{eq:conditionlambda0}
    F_{ik}\tilde\lambda_{kj}+F_{kj}\tilde\lambda_{ki}=0\,.
\end{equation}
This equations states that $\tilde\lambda_{ij}$ corresponds to a rotation symmetry acting on the $x^i$ for every rotation that leaves $F_{ij}$ invariant.
The other gauge transformation parameters are given by
\begin{eqnarray}
    \Omega & = & 0\,,\\
    \tilde\Omega & = & -\frac{1}{2}\left(b+ct\right)\,,\\
    \lambda & = & -e-\frac{c}{2}v\,,\\
    \lambda^0_i & = & -\partial_t a^i+\frac{c}{4}x^i\,,\\
    \lambda^1_i & = & \partial_v a^i-\frac{1}{2}F_{ij}\left(e+\frac{c}{2}v\right)x^j\,,\\
    \Sigma_t & = & \partial_t\Sigma-\partial_t\tilde\Sigma+x^i\partial_v a^i\,,\\
    \Sigma_v & = & \partial_v\Sigma-\partial_v\tilde\Sigma+\frac{1}{2}F_{ij}x^i a^j\,,\\
    \Sigma_i & = & \partial_i\Sigma\,,
\end{eqnarray}
where $\Sigma$ and $\tilde\Sigma$ are arbitrary functions on spacetime. We note that the 1-form gauge transformation parameter $\Sigma_M$ is itself only determined up to a gauge transformation of the form $\delta\Sigma_M=\partial_M\Sigma$.
Furthermore, we note that the $z=1$ Weyl transformations corresponding to $\Omega$ do not play a role in the symmetries of these backgrounds.
Instead, the emergent $z=2$ Weyl transformations corresponding to $\tilde\Omega$ appear.

For the embedding fields $X^M$ the global symmetries take the form
$\delta X^M=-\xi^M$, where we replaced target space coordinates with embedding fields,
\begin{eqnarray}
    \xi^t & = & a+bX^t+\frac{c}{2}(X^t)^2\,,\\
    \xi^v & = & d+eX^t+\frac{X^v}{2}(b+cX^t)\,,\\
    \xi^i & = & a^i(X^t,X^v)+\lambda^i{}_j X^j-\frac{1}{4}\left(b+cX^t\right)X^i\,.
\end{eqnarray}
More generally, one can combine these with local worldsheet symmetries such that
$\delta X^M=\zeta^\alpha\partial_\alpha X^M-\xi^M$.

In the static gauge \eqref{eq:staticgauge} we require that $\delta X^t=0$ and $\delta X^v=\delta f$. This gives us 
\begin{eqnarray}
    \zeta^\tau & = & a^\tau+2s\tau+\frac{k}{2}\tau^2\,,\label{eq:zetatau}\\
    \zeta^\sigma & = & a^\sigma+\tilde v\tau+\left(s+\frac{k}{2}\tau\right)\sigma\,,\label{eq:zetasigma}
\end{eqnarray}
where we defined
\begin{eqnarray}
    a^\tau & = & \frac{a}{K}\,,\\
    s & = & \frac{b}{2} \,,\\
    k & = & cK\,,\\
    a^\sigma & = & \frac{d}{w R_v}\,,\\
    \tilde v & = & \frac{Ke}{wR_v}\,.
\end{eqnarray}
In static gauge, the global symmetries are realized on the $X^i=X^i(\tau, \sigma)$ and $f(\tau)$ as 
\begin{eqnarray}
    \delta X^i & = & \zeta^\tau\partial_\tau X^i+\zeta^\sigma\partial_\sigma X^i-\xi^i\,,\label{eq:sym1}\\
    \delta f & = & \zeta^\tau\dot f-\frac{1}{2}\partial_\tau\zeta^\tau f\,,\label{eq:sym2}
\end{eqnarray}
where $\zeta^\tau$ and $\zeta^\sigma$ are given by \eqref{eq:zetatau} and \eqref{eq:zetasigma}, and where $\xi^i$ is given by
\begin{equation}
    \xi^i = \tilde a^i(\tau,\sigma)+\lambda^i{}_j X^j-\frac{1}{2}\left(s+\frac{k}{2}\tau\right)X^i\,.
\end{equation}
Here, $a^i(X^t, X^v)=\tilde a^i(\tau,\sigma)$ which obeys
\begin{equation}\label{eq:tildeai}
    F_{ij}\left(u\partial_\tau\tilde a^i-T_g\dot f \partial_\sigma \tilde a^i\right)+\tilde u^2\partial_\sigma^2 \tilde a^j=0\,,
\end{equation}
as follows from \eqref{eq:conditionai}, and where $\lambda_{ij}$ is given by
\begin{equation}\label{eq:lambdaij}
     \lambda_{ij}=\tilde\lambda_{ij}-\frac{u}{2\tilde u^2}F_{ij}\left(\sigma+\frac{f}{wR_v}\right)\left(\tilde v+\frac{1}{4}k\left(\sigma+\frac{f}{wR_v}\right)\right)\,.
\end{equation}
We note that \eqref{eq:tildeai} is the same equation as~\eqref{eq:EOMXi}, the equation of motion of the $X^i$.

To summarize, the symmetries of the Galilean string in these backgrounds (and their parameters) are as follows: time translations ($a^\tau$), $z=2$ scale transformations ($s$), special conformal transformations ($k$), longitudinal translations ($a^\sigma$), and Galilean boost~($\tilde v$).
We furthermore have rotations ($\tilde\lambda_{ij}$) that leave invariant the symplectic 2-form $F_{ij}$.
They include as a special case $\tilde\lambda_{ij}=\alpha F_{ij}$ for any real $\alpha$.
The transformations whose parameters are $a^\tau, s, k, a^\sigma, \tilde v, \alpha$ form the two-dimensional Schr\"odinger algebra.
Finally, any $\tilde a^i(\tau,\sigma)$ that solves \eqref{eq:tildeai} is a symmetry as well. 

\subsection{Algebra of Noether charges}
Next, we will determine the Noether charges for the global symmetries of the gauge fixed theory. The transformations \eqref{eq:sym1} and \eqref{eq:sym2} transform the Lagrangian \eqref{eq:toymodel} as follows,
\begin{equation}\label{eq:variationtoymodel}
    \delta\mathcal{L}=\partial_\alpha\left(\zeta^\alpha\mathcal{L}+K^\alpha\right)\,,
\end{equation}
where $K^\alpha$ is given by
\begin{eqnarray}
    K^\tau & = & \frac{u}{2}F_{ij}a^i X^j\,,\\
    K^\sigma & = & \tilde u^2 X^i\partial_\sigma a^i-\frac{1}{2}T_g\dot f F_{ij}a^i X^j\,.
\end{eqnarray}
The vector $K^\alpha$ only depends on $a^i$, the inhomogeneous part of $\delta X^i$. Furthermore, the vector $K^\alpha$ describes the extent to which the Lagrangian density does not transform as a scalar density of weight one.
Put differently, the $K^\alpha$ term makes it so that the action is invariant up to possibly a boundary term. Such symmetries are often associated with central extensions of the algebra described by the Lie brackets between the vectors~$\xi^M\partial_M$. We will show that this is indeed the case here, too. To this end we need to work out the Poisson algebra of the Noether charges on phase space.

Applying Noether's theorem to \eqref{eq:variationtoymodel} we obtain the Noether current\footnote{We note that in \eqref{eq:toymodel} the field $f(\tau)$ is not a 2-dimensional variable. We can make it one by adding a Lagrange multiplier term, $\chi\partial_\sigma f(\tau,\sigma)$ to \eqref{eq:toymodel}, where $\chi$ is the Lagrange multiplier field that is periodic in $\sigma$. It is in this context that we apply the usual 2-dimensional Noether theorem. The equation of motion of $f$ then involves $\partial_\sigma\chi$. After integrating this over $\sigma$, and using the periodicity of $\chi$, we obtain the 1-dimensional equation of $f(\tau)$ which takes the form of a conserved charge $\frac{d}{d\tau}\oint d\sigma m_i X'^i=0$.\label{fnote:EOMf}} 
\begin{equation}
    J^\alpha=\frac{\partial\mathcal{L}}{\partial\partial_\alpha X^i}\delta X^i+\frac{\partial\mathcal{L}}{\partial\partial_\alpha f}\delta f-\zeta^\alpha\mathcal{L}-K^\alpha\,,
\end{equation}
where $\delta X^i$ and $\delta f$ are given in \eqref{eq:sym1} and \eqref{eq:sym2}. The Noether charge $Q[\zeta^\tau, \zeta^\sigma, a^i]$ is given by
\begin{equation}\label{eq:Noethercharge}
    Q[\zeta^\alpha, \tilde\lambda_{ij}, \tilde a^i]=\oint d\sigma \left[\zeta^\tau \frac{\tilde u^2}{2}X'^i X'^i-\zeta^\sigma u m_i X'^i+u m_i\left(\lambda^i{}_jX^j+2\tilde a^i\right)-\frac{1}{2}T_g f m_i X'^i\partial_\tau\zeta^\tau\right]\,,
\end{equation}
where $\zeta^\tau$, $\zeta^\sigma$, $\lambda_{ij}$ and $\tilde a^i$ are given in Equations \eqref{eq:zetatau}, \eqref{eq:zetasigma}, \eqref{eq:lambdaij}, and \eqref{eq:tildeai}, respectively. More explicitly, we have the following Noether charges,
\begin{eqnarray}
    Q[a^\tau] & = & a^\tau \oint d\sigma \left[\frac{\tilde u^2}{2}X'^i X'^i\right]\,,\\
    Q[s] & = & s\oint d\sigma \left[\tau \tilde u^2 X'^i X'^i+\frac{u}{2}\left(\sigma+\frac{f}{wR_v}\right) F_{ij} X'^i X^j\right]\,,\\
     Q[k] & = & k\oint d\sigma \left[\tau^2 \frac{\tilde u^2}{4}X'^i X'^i+\frac{u}{4}\tau\left(\sigma+\frac{f}{wR_v}\right)  F_{ij} X'^i X^j\right.\nonumber\\
     &&\left.+\frac{T_g u}{16\tilde u^2}\left(\sigma+\frac{f}{wR_v}\right)^2F_{ij}F_{ik} X^j X^k\right]\,,\\
     Q[a^\sigma] & = & a^\sigma\oint d\sigma  \frac{u}{2} F_{ij} X'^iX^j\,,\\
     Q[\tilde v] & = & \tilde v\oint d\sigma \left[\frac{u}{2}\tau F_{ij} X'^i X^j+\frac{u^2}{4\tilde u^2} 
    \left(\sigma+\frac{f}{wR_v}\right)F_{ij}F_{ik}X^jX^k\right]\,,\\
    Q[\alpha] & = & -\alpha\oint d\sigma \left[\frac{u}{2} F_{ij}F_{ik}X^jX^k\right]\,,\\
     Q[\tilde\lambda] & = & \frac{1}{2}\tilde\lambda_{ij}\oint d\sigma \frac{u}{2}\left[ X^i F_{jk} X^k-X^j F_{ik} X^k\right]\,,\\
     Q[\tilde a] & = & \oint d\sigma u F_{ij}X^i \tilde a^j(\tau,\sigma)\,,
\end{eqnarray}
where $\tilde\lambda_{ij}$ obeys \eqref{eq:conditionlambda0} and is not proportional to $F_{ij}$. 

The Poisson bracket \eqref{eq:Poissonbracket} tells us that
\begin{equation}
    \{Q[\tilde a]\,,Q[\tilde b]\}=\oint d\sigma u F_{ij}\tilde a^i\tilde b^j\,.
\end{equation}
This leads to a central extension of the algebra that we get from the Lie brackets between the vector fields $\xi^M\partial_M$.
We can get the full algebra from the following result
\begin{equation}
    \{Q[\zeta_1^\alpha, \tilde\lambda_{1ij}, \tilde a_1^i]\,,Q[\zeta_2^\alpha, \tilde\lambda_{2ij}, \tilde a_2^i]\}=Q[\zeta_3^\alpha, \tilde\lambda_{3ij}, \tilde a_3^i]+\oint d\sigma u F_{ij}\tilde a_1^i\tilde a_2^j\,,
\end{equation}
where
\begin{eqnarray}
    \zeta_3^\alpha & = & [\zeta_1\,,\zeta_2]^\alpha\,,\\
    \tilde a_3^i & = & \zeta^\alpha_1\partial_\alpha\tilde a^i_2-\lambda_1^i{}_j\tilde a^j_2+\frac{1}{2}\partial_\sigma\zeta^\sigma_1 \tilde a^j_2-\frac{1}{wR_v}\zeta^\tau_1\dot f\partial_\sigma\tilde a^i_2\nonumber\\
    &&+\frac{f}{wR_v}\partial_\sigma\zeta^\sigma_1\partial_\sigma\tilde a^i_2-\left(1\leftrightarrow 2\right)\,.
\end{eqnarray}
If we parameterize $\zeta_3^\alpha$ in the same way as we did for $\zeta_1^\alpha$ and $\zeta_2^\alpha$, i.e. as
\begin{eqnarray}
    \zeta_3^\tau & = & a_3^\tau+2s_3\tau+\frac{k_3}{2}\tau^2\,,\\
    \zeta_3^\sigma & = & a_3^\sigma+\tilde v_3\tau+\left(s_3+\frac{k_3}{2}\tau\right)\sigma\,,
\end{eqnarray}
then we can write $\lambda_{3ij}$ in the expression for $Q[\zeta_3^\alpha, \tilde\lambda_{3ij}, \tilde a_3^i]$ as follows,
\begin{equation}
     \lambda_{3ij}=\alpha_3F_{ij}-\frac{u}{2\tilde u^2}F_{ij}\left(\sigma+\frac{f}{wR_v}\right)\left(\tilde v_3+\frac{1}{4}k_3\left(\sigma+\frac{f}{wR_v}\right)\right)\,,
\end{equation}
where 
\begin{equation}
    \alpha_3=-\frac{u}{2\tilde u^2}\left(a^\sigma_1\tilde v_2-a^\sigma_2\tilde v_1\right)\,.
\end{equation}
In other words,
$\tilde\lambda_{3ij}=\alpha_3 F_{ij}\,$.

So far, we have only looked at the situation locally. A true symmetry respects the boundary conditions imposed on the fields which in this case are periodic boundary conditions for the $X^i$ due to the fact that we are on a cylinder. The identification $\sigma\sim\sigma+2\pi$ breaks the transformations with parameters $s, k, \tilde v$ and requires that the $\tilde a^i$ are periodic in $\sigma$.

\section{Discussion}
\label{sec:concl}

In this work, starting from non-relativistic string theory coupled to torsional string Newton--Cartan geometry, we have considered two further limits, leading
to two new types of worldsheet theories, exhibiting Galilean and Carrollian worldsheet structures, respectively. 
We have furthermore considered various aspects of gauge fixings of these actions.
Finally, for the Galilean string, we have studied in detail
a solvable model related to a class of target spacetimes that arise in the putative holographically dual description of Spin Matrix theories.\footnote{See \cite{Harmark:2019zkn,Baiguera:2020jgy,Baiguera:2020mgk,Baiguera:2021hky,Baiguera:2022pll} for recent work on the dual NR field theories.}
For this solvable model we have discussed the phase space, mode expansion, symmetries and algebra of Noether charges.

Clearly, it would be important to consider the quantization of these new worldsheet theories.
As a first check, it would then be interesting to see if reproducing the global symmetries that we obtained for the classical theory from the quantum theory imposes particular restrictions.
In addition, it would be useful to find a class of solvable models for the Carrollian string,
in analogy with the Galilean solvable model presented in this paper that was motivated from Spin Matrix limits of AdS/CFT.
For the Carrollian string, we currently do not have such a shortcut, so a careful analysis would require a better understanding of consistent backgrounds from the appropriate quantum consistency conditions.
Furthermore, we have not taken into account the dilaton term in the present work (but see earlier ideas in~\cite{Harmark:2019upf}).
We hope to return to these aspects in future work. 

Relatedly, it would be very interesting to find out if the Carrollian limit of this paper has any realization in the context of the AdS/CFT correspondence. 
This could open various interesting holographic applications of the Carrollian string that we have found.
It would also be interesting to study possible connections to other work on Carroll string sigma models including~\cite{Isberg:1993av,Bagchi:2013bga,Bagchi:2016yyf,Bagchi:2020fpr,Bagchi:2021ban}.

An obvious but important generalization of the present work is to obtain the supersymmetric version%
\footnote{See \cite{Blair:2020gng} for a worldsheet action of NR superstrings obtained via a connection to double
field theory and \cite{Bergshoeff:2021tfn,Bergshoeff:2022pzk} for actions  and solutions of NR supergravity in ten dimensions
as well as the review \cite{Bergshoeff:2022iyb} on non-Lorentzian supergravity.} 
 of the TSNC strings and the two limits we have obtained. This is interesting in its own right, and in addition expected to be important for applications in holography. 

Additionally, the Galilean and Carrollian worldsheet limits we considered in this paper each keep the leading order terms on the worldsheet.
One could alternatively consider a limit where the leading order term is cancelled using a Lagrange multiplier, which focuses on the dynamics of the next-to-leading-order term.
(In the Carroll case, such limits are conventionally referred to as `magnetic', while the former are `electric', and vice versa for the Galilean case.
See for example~\cite{Bellac-Levy-Leblond,Duval:2014uoa,Festuccia:2016caf,Henneaux:2021yzg,deBoer:2021jej,Hansen:2021fxi,Baiguera:2022lsw,Campoleoni:2022ebj,deBoer:2023fnj}.)
It would be interesting to see if such limits also give rich actions in the present case.%
\footnote{
    We thank the anonymous referee for raising this point.
}

Another direction that would be worthwhile to pursue is to consider expansions as opposed to limits with respect to the longitudinal speed of light $\tilde c$. 
We note that the large (transverse) speed of light expansion of string theory%
\footnote{This development built on earlier work that developed the large speed of light expansion of GR \cite{VandenBleeken:2017rij,Hansen:2019pkl,Hansen:2020pqs}, see \cite{Hartong:2022lsy} for a review. See also \cite{Hansen:2021fxi}  for the small speed of light expansion of GR.} 
was considered in \cite{Hartong:2021ekg,Hartong:2022dsx}. Within this framework, the TSNC string follows from a certain limiting procedure.
Likewise, in view of the results of this paper, it could be interesting to consider 
a large $\tilde c$ expansion as opposed to the $\tilde c \rightarrow \infty$ 
  limit considered in this paper. From the dual SMT perspective this would correspond to a small $\lambda$ expansion on the field theory side \cite{Harmark:2008gm}. Similarly, the small $\tilde c$ expansion could be interesting to consider. 

  From a more general perspective, one could also examine whether  Galilean/Carrollian limits of the type discussed here can be applied to $p$-brane actions (see e.g. \cite{Bergshoeff:2020xhv}).
In particular, it would be interesting to  explore its relation to non-relativistic D-brane theories \cite{Gomis:2020fui,Gomis:2020izd,Ebert:2021mfu}
as well as non-relativistic M-theory and M-branes  \cite{Blair:2021waq,Kluson:2021djs,Kluson:2021pux,Roychowdhury:2022est,Ebert:2023hba}.

\section*{Acknowledgements}

We thank Eric Bergshoeff, Andrea Fontanella and Ziqi Yan for useful discussions. The work of JH is supported by the Royal Society University Research Fellowship (renewal) ``Non-Lorentzian String Theory'' (grant number URF\textbackslash R\textbackslash221038)
and in part by the Leverhulme Trust Research Project Grant (RPG-2019-218) “What is Non-Relativistic Quantum Gravity and is it Holographic?”. The work of LB was supported by the Royal Society Research Fellows Enhancement Award 2017 “Non-Relativistic Holographic Dualities” (grant number RGF$\backslash$EA$\backslash$180149). The work of TH, NO and GO is supported in part by the project ``Towards a deeper understanding of  black holes with non-relativistic holography'' of the Independent Research Fund Denmark (grant number DFF-6108-00340).
The work of NO and GO is supported in part by VR project grant 2021-04013. The work of NO is also supported by the Villum Foundation Experiment project 00050317, ``Exploring the wonderland of Carrollian physics: Extreme gravity, spacetime horizons and supersonic fluids''. Nordita is supported in part by Nordforsk.

\appendix

\section{Gauge symmetry algebras and target spacetime geometry \label{app:SymAlg} }

In this appendix we present the underlying gauge symmetry algebras of the target spacetime geometries that appear in the worldsheet actions of the Galilean and Carrollian longitudinal limits of the NR string that we derived in Section~\ref{sec:limits}.

As a starting point, we recall that it was shown in \cite{Bidussi:2021ujm} that the target space geometry of the NR string arises from the gauging of a novel algebra, the so-called F-string Galilei (FSG) algebra.
This algebra is obtained by an İnönü–Wigner (IW) contraction of the Poincaré algebra extended by the symmetries of the Kalb–Ramond field.   We refer to~\cite{Bidussi:2021ujm} for details on this algebra and its contraction to FSG, as well as the associated gauging procedure, summarizing here the main results. 

Using $A=0,1$ and $a=2 , \ldots , d+1$ for the longitudinal and transverse tangent space indices
respectively, the FSG algebra has the following generators:
longitudinal Lorentz boosts $J_{AB} = \epsilon_{AB} J$,
transverse rotations $J_{ab} $,
stringy Galilean boosts $G_{Ab}$,
longitudinal and transverse translations $H_{A}$ and $P_a$,
as well as longitudinal and transverse generators $N_A$ and $Q_a$ associated with one-form gauge symmetry of the TSNC two-form $m_{\mu\nu}$. 

The non-zero commutators of the algebra are given by the boost relations
\begin{subequations}
\label{FSG_algebra}
  \begin{align}
    [G_{Ab}, H_C]
    &= \eta_{AC} P_b + \epsilon_{AC} Q_b\, ,
    \label{GH}
    \\
    [G_{Ab}, P_c]
    &= - \delta_{bc} \epsilon_A{}^B N_B\, ,
    \label{GPt}
    \\
    [G_{Ab}, Q_c]
    &= - \delta_{bc} N_A\, ,
    \label{GQ}
  \end{align}
\end{subequations}
where we note in particular that  $[G_{Ab},G_{Cd}]=0$.
In addition, we have the commutators that express the behaviour under transverse rotations and longitudinal Lorentz boosts, which take the form 
\begin{subequations}
\label{boring_algebra}
  \begin{align}
    [J_{ab}, J_{cd}]
    &= \delta_{ac} J_{bd}
    - \delta_{bc} J_{ad}
    + \delta_{bd} J_{ac}
    - \delta_{ad} J_{bc}\, ,
    \\
    [J_{ab}, G_{Cd}]
    &= \delta_{ad} G_{Cb} - \delta_{bd} G_{Ca}\, ,
    \\
    [J_{ab},P_c]
    &= \delta_{ac} P_b - \delta_{bc} P_a\, ,
    \\
    [J_{ab},Q_c]
    &= \delta_{ac} Q_b - \delta_{bc} Q_a\, ,
    \\
    [J,G_{Ab}]
    &= \epsilon^C{}_A G_{Cb}\, ,
    \\
    [J, H_A]
    &= \epsilon^B{}_A H_B\, ,
    \\
    [J, N_A]
    &= \epsilon^B{}_A N_B\, .
  \end{align}
\end{subequations}
We use the conventions $\epsilon^{01} = -\epsilon_{01} = 1$.

The relation between the algebra above and the geometric fields in the TSNC geometry is encoded in the following the Lie algebra-valued connection
\begin{equation}
  \label{gaugeconnection}
  \mathcal{A}_M
  = \tau_M^A H_A + e_M^a P_a + \pi_M^A N_A + \pi_M^a Q_a + \cdots \,,
\end{equation}
where $\cdots$ represents further terms involving spin connection-type fields. 
The TSNC background geometry then consists of 
the longitudinal vielbeine $\tau_M^A$ and the combinations
\begin{equation}
  \label{eq:tsnc-composite-objects}
  h_{MN}
  =  e_M^a e_N^b \delta_{ab}\,,
  \qquad
  m_{MN}
  = \tau_{[M}^A\pi_{N]}^B \eta_{AB} +e_{[M}^a \pi_{N]}^b \delta_{ab}\,.
\end{equation}
The symmetries of this geometry, including diffeomorphisms and one-form  gauge symmetries, can then be found from an appropriate gauging prescription of the FSG algebra, as shown in detail in~\cite{Bidussi:2021ujm}.
The worldsheet action~\eqref{NG_TSNC} of the TSNC string is invariant under the resulting symmetries. 

Similarly, one can ask what the corresponding underlying symmetry algebra is for the target space geometries
that appear in the Galilean and Carrollian longitudinal limits of the NR string obtained in Sections~\ref{sec:Gal_long_limit} and \ref{sec:Car_long_limit} of the main text.
We present these algebras below, though we omit a detailed exposition of how their gauging gives rise to the corresponding symmetries (notably the longitudinal and transverse local boost transformations), which follows along the lines of Section~2 of~\cite{Bidussi:2021ujm}.
The relevant algebras follow from the FSG algebra by introducing the longitudinal speed of light $\tilde c$ and subsequently taking an IW contraction by sending $\tilde c$ to $\infty$ and $0$ respectively. 

\subsubsection*{Symmetry algebra of Galilean string}
Translating the scaling of the TSNC fields and the symmetry transformations that we obtained in Section \ref{sec:Gal_long_limit} to the algebra using~\eqref{gaugeconnection} and~\eqref{eq:tsnc-composite-objects}, we see that
the longitudinal Galilean limit corresponds to scaling the generators as follows,
\begin{equation}
\begin{array}{c}
  H_0 = \tilde{H}_0
  \spa
  H_1 = \tilde{c}\,  \tilde{H}_1
  \spa
  P_a = \tilde{P}_a
  \spa
  N_0 = \frac{1}{\tilde c}\, \tilde{N}_0
  \spa 
  N_1 = \frac{1}{\tilde c^2}\, \tilde{N}_1
  \spa
  Q_a = \frac{1}{\tilde c}\, \tilde{Q}_a,
  \\[2mm]
  G_{0a} =  \tilde{G}_{0a}
  \spa
  G_{1a} = \frac{1}{\tilde c}\, \tilde{G}_{1a}
  \spa
  J= \tilde{c}\, \tilde{J}
  \spa
  J_{ab} = \tilde{J}_{ab} \,,
\end{array}
\end{equation}
and subsequently taking the IW contraction $\tilde c \rightarrow \infty$.

After removing the tildes, the resulting algebra has the following non-zero commutation relations. 
The commutators \eqref{FSG_algebra} involving string Galilean boosts become
\begin{subequations}
\label{GCAFSG_algebra}
  \begin{align}
    [G_{0a}, H_0]
    &= - P_a\,,
    \\
    [G_{1a}, H_1]
    &= P_a\,,
    \\
    [G_{1a}, H_0]
    &= Q_a\,,
    \\
    [G_{1a}, P_b]
    &= \delta_{ab}  N_0\,,
    \\
    [G_{Ab}, Q_c]
    &= - \delta_{bc} N_A\,.
  \end{align}
\end{subequations}
The commutators in \eqref{boring_algebra} involving the longitudinal Lorentz boosts $J$ become
\begin{subequations}
\label{boring_algebra_Gal}
  \begin{align}
    [J,G_{1a}]
    & = G_{0a}\,,
    \\
    [J, H_0]
    & = H_1\,,
    \label{longGal}
    \\
    [J,N_1]
    & = N_0\,,
  \end{align}
\end{subequations}
while all commutators involving $J_{ab}$ remain unchanged, which reflects that we still have $SO(d-1)$
symmetry in the transverse directions.
Importantly, we recognize from the commutator  \eqref{longGal} that the generator $J$ has become
a Galilean boost in the longitudinal subspace. 
Using the gauging methods described in e.g. \cite{Hartong:2015zia} and \cite{Bidussi:2021ujm} it is then
not difficult to show that the longitudinal and transverse boost symmetry~\eqref{lGlb} and \eqref{stringGal_GCA} of the action~\eqref{GCA_NG} directly follow from the algebra above. 

More generally, we thus conclude that the  local symmetry
of the target spacetime in the longitudinal Galilean limit is described by the particular 
 contraction of the FSG algebra given above. 
For a subclass of spacetimes, the resulting geometry reduces to the $U(1)$-Galilean geometries that were considered previously in~\cite{Harmark:2017rpg,Harmark:2018cdl,Harmark:2019upf,Harmark:2020vll}. 

\subsubsection*{Symmetry algebra of Carrollian string}
Likewise, we can translate the limit that we described in Section~\ref{sec:Car_long_limit} on the level of the TSNC fields and the symmetry parameters to the FSG algebra using~\eqref{gaugeconnection} and~\eqref{eq:tsnc-composite-objects},
\begin{equation}
  \begin{array}{c}
    H_0 = \tilde{H}_0
    \spa
    H_1 = \tilde{c}\, \tilde{H}_1
    \spa
    P_a = \tilde{P}_a
    \spa
    N_0 = \tilde{c}\, \tilde{N}_0
    \spa 
    N_1 = \tilde{N}_1
    \spa
    Q_a = \tilde{c}\, \tilde{Q}_a 
    \\[2mm]
    G_{0a} = \tilde{G}_{0a}
    \spa
    G_{1a} = \frac{1}{\tilde{c}}\, \tilde{G}_{1a}
    \spa
    J = \frac{1}{\tilde{c}} \tilde{J}
    \spa
    J_{ab} = \tilde{J}_{ab} \,,
  \end{array}
\end{equation}
followed by the IW contraction $\tilde c \rightarrow 0$. 

After removing the tildes, this leads to the following non-zero commutators.
The commutators for the longitudinal rotations $J_{ab}$ again remain unchanged, while
the string Galilean boost commutators become
\begin{subequations}
\label{BMSFSG_algebra}
  \begin{align}
    [G_{0a}, H_0]
    &= - P_a\,,
    \\
    [G_{1a}, H_1]
    &= P_a\,,
    \\
    [G_{0a}, H_1]
    &= - Q_a \,,
    \\
    [G_{0a}, P_b]
    &= \delta_{ab}  N_1 \,,
    \\
    [G_{Ab}, Q_c]
    &= - \delta_{bc} N_A \,.
  \end{align}
\end{subequations}
Finally, the longitudinal Lorentz boosts~\eqref{boring_algebra} now become
\begin{subequations}
\label{boring_algebra_Car}
  \begin{align}
    [J,G_{0a}]
    &= G_{1a}\,,
    \\
    [J, H_1]
    &= H_0\,,
    \label{longCar}
    \\
    [J, N_0]
    &= N_1 \,.
  \end{align}
\end{subequations}
This time,
as evidenced by~\eqref{longCar},
the longitudinal Lorentz boost becomes a Carroll boost.
Again, this algebra can be gauged to obtain the target space symmetry transformation of the action~\eqref{BMS_NG} that follow from a limit of the TSNC transformations.

\bibliographystyle{JHEP}
\bibliography{biblio}

\end{document}